\documentclass[preprint]{aastex}

\slugcomment{Accepted for publication in AJ, 16 March 2008}

\shorttitle{GO-10775 Reductions}
\shortauthors{Anderson}

\def\secspt{$\buildrel{\prime\prime}\over .$}

\begin{document}

\title{An ACS Survey of Globular Clusters V:  Generating a Comprehensive
       Star Catalog for Each Cluster\footnote{
               Based on observations with the NASA/ESA 
               {\it Hubble Space Telescope}, obtained at the 
               Space Telescope Science Institute, which is operated by 
               AURA, Inc., under NASA contract NAS 5-26555.}} 

\author{Jay Anderson}
\affil{Space Telescope Science Institute,
       Baltimore, MD 21218, USA; jayander@stsci.edu}

\author{Ata Sarajedini}
\affil{Department of Astronomy, University of Florida, 
       Gainesville, FL 32611, USA; ata@astro.ufl.edu}

\author{Luigi R. Bedin}
\affil{Space Telescope Science Institute, 
       Baltimore MD 28218, USA; bedin@stsci.edu}

\author{Ivan R. King}
\affil{Department of Astronomy, University of Washington, 
       Seattle, WA 98195-1580, USA; king@astro.washington.edu}

\author{Giampaolo Piotto}
\affil{Dipartimento di Astronomia, Universit\`a di Padova, 
       35122 Padova, Italy; giampaolo.piotto@unipd.it}

\author{I. Neill Reid}
\affil{Space Telescope Science Institute, 
       Baltimore MD 28218, USA; inr@stsci.edu}

\author{Michael Siegel}
\affil{University of Texas, McDonald Observatory, 
       Austin, TX 78712, USA; siegel@astro.as.utexas.edu}

\author{Steven R. Majewski}
\affil{Department of Astronomy, University of Virginia, 
       Charlottesville, VA 22904-4325, USA; srm4n@virginia.edu}

\author{Nathaniel E. Q. Paust}
\affil{Space Telescope Science Institute, 
       Baltimore MD 28218, USA; npaust@stsci.edu}

\author{Antonio Aparicio}
\affil{Instituto de Astrof\'isica de Canarias, V\'ia L\'actea s/n,
       E-38200 La Laguna, Spain; antapaj@iac.es}

\author{Antonino P. Milone}
\affil{Dipartimento di Astronomia, Universit\`a di Padova, 
       35122 Padova, Italy; antonino.milone@unipd.it}

\author{Brian Chaboyer}
\affil{Department of Physics and Astronomy, Dartmouth College, 
       Hanover, NH 03755, USA; chaboyer@heather.dartmouth.edu}

\author{Alfred Rosenberg}
\affil{Instituto de Astrof\'isica de Canarias, 
       E-38200 La Laguna, Canary Islands, Spain; alf@iac.es}

\affil{March 16, 2008}

\begin{abstract}
The ACS Survey of Globular Clusters has used HST's Wide-Field Channel to
obtain uniform imaging of 65 of the nearest globular clusters to provide
an extensive homogeneous dataset for a broad range of scientific
investigations.  The survey goals required not only a uniform observing
strategy, but also a uniform reduction strategy.  To this end, we
designed a sophisticated software program to process the cluster data in
an automated way.  The program identifies stars simultaneously in the
multiple dithered exposures for each cluster and measures them using the
best available PSF models.  We describe here in detail the program's
rationale, algorithms, and output.  The routine was also designed to
perform artificial-star tests, and we run a standard set of $\sim$10$^5$
tests for each cluster in the survey.  The catalog described here will
be exploited in a number of upcoming papers and will eventually be made
available to the public via the world-wide web.
\end{abstract}



\section{INTRODUCTION}
\label{s.INTRO}
The Galaxy's globular clusters hold important clues to a large number of
scientific questions, ranging from star formation to stellar structure,
galaxy evolution, and cosmology.  Many of these questions can be
answered only by surveying a significant fraction of the clusters and
studying the cluster system as an ensemble.  Initial globular cluster
surveys (e.g., Zinn 1980, Armandroff 1989) focused on integrated-light
properties such as total brightness, colors, metallicity, and reddening.
Many subsequent ``surveys'' have been constructed by assembling various
data from the multitude of independent observations of individual
clusters (e.g., Djorgovski \& King 1986, Djorgovski \& Meylan 1993, 
Trager et al.\ 1995, Lee et al.\ 1996, and Harris 1996).  However, since 
each cluster is typically observed with a different instrument and under 
different conditions, there are limits to how homogeneous such a 
patched-together data set can be.

In an effort to construct a more homogeneous sample, Rosenberg 
et al.\ (2000a, 2000b) surveyed 56 clusters from the ground, producing 
star catalogs and color-magnitude diagrams (CMDs) that can be directly 
intercompared to yield relative ages and relative horizontal-branch 
morphologies.  Piotto et al.\ (2002) used WFPC2 snapshots to image
the central regions of 74 clusters and construct CMDs in a uniform
photometric system.  These surveys have allowed clusters to be studied 
on a more even footing than ever before, but the data in these surveys
still suffer from severe crowding in the cluster cores, irregularities 
in the sampling, and gaps in the field of view.  Thanks to its fine 
sampling, large dynamic range and wide, contiguous field of view, the 
Advanced Camera for Survey's (ACS's) Wide-Field Channel (WFC) on board 
the Hubble Space Telescope (HST) is the first instrument that can 
improve dramatically on all of these shortcomings.  

The ACS Survey of Globular Clusters presented here was designed to
provide a nearly complete catalog of all the stars present in the
central two arcminutes of 65 targeted clusters.  Such a uniform data set
has many scientific applications, and we are currently in the process of
using the catalog for broad studies of:\ binary-star distributions,
absolute and relative ages, horizontal-branch morphology, blue
stragglers, isochrone fitting, mass functions, and dynamical models.  We
are also measuring internal motions and orbits for those clusters that have
sufficient archival data.  In addition to addressing these major
scientific issues, one of the main legacies of this survey will be to
provide the community with a definitive catalog of stars in the central
regions of these clusters.  This data base will serve as a touchstone
for studies of these clusters for many years to come, and as such it
should be as accurate and comprehensive as possible.

The images that make up this survey consist almost entirely of point 
sources, but each cluster has a different central concentration and density 
profile.  So, to construct a definitive catalog, we needed a star-finding 
and measuring routine that works in a variety of crowding situations, 
often across the same cluster field.  With this in mind, we developed a 
sophisticated computer program that simultaneously analyzes all of the 
survey exposures for each cluster (one short exposure plus four to five 
deep exposures for each of the F606W and F814W filters), to construct a 
single list of detected stars and their measured parameters.  The routine 
was designed to deal well with both crowded and uncrowded situations, 
and as such it is able to find almost every star that a human could find. 
At the same time, the routine uses the independence of the pointings and 
knowledge of the PSF to avoid including image artifacts in the list.  

This paper describes the data-reduction procedure we developed and the 
resulting catalog we produced for each cluster.  It is organized as 
follows:  We begin by describing the observations we have available for 
each cluster (\S\ \ref{s.OBSNS}) and the preliminary set-up steps required 
before the finding program could be run on the images (\S\ \ref{s.SETUP}).  
Before diving into the details of our procedures, we first give an overview 
of the general considerations that are involved in finding and measuring 
stars in dithered, undersampled images of globular clusters 
(\S\ \ref{s.OVERVIEW}).  We then describe in detail our automated finding 
and measuring program and use it to construct a catalog of the real 
stars for each cluster (\S\ \ref{s.KSYNC}).  We use the same program 
to perform a standard battery of artificial-star tests for each cluster 
(\S\ \ref{s.ASTEST}).  We also consider the photometric errors that are 
present in an ACS data set such as that collected here 
(\S\ \ref{s.PHOTO_ERRORS}).  Finally, we describe the photometric and 
astrometric calibration and the assembly of the final catalog of positions, 
magnitudes, quality characterizations, etc., for the detected stars 
for each cluster field (\S\ \ref{s.CATALOG}).  We end with a summary of 
upcoming scientific results and additional studies that will complement 
this survey (\S\ \ref{s.SUMMARY}).

%

\section{OBSERVATIONS}
\label{s.OBSNS}

The goal of the ACS Survey of Globular Clusters (GO-10775, PI-Sarajedini)
was to image the central regions of a large number of globular clusters 
in order to generate a homogeneous set of star catalogs.  The clusters 
are all at different distances and all have different central densities and 
radial profiles, so there is of course no way to obtain identical data 
for every cluster, but our aim was to come as close to this ideal as 
possible.

Each cluster was observed for one orbit in F606W ($V$) and one orbit in 
F814W ($I$), except for M54, which was observed for 2 orbits in each filter.
In each orbit, we took one short exposure and either four or five deeper 
exposures, depending on how many we could fit into the orbit.  We chose the 
exposure times for each cluster so that the horizontal-branch stars would 
be unsaturated in the short exposure and the turn-off and subgiant branch 
stars would be unsaturated in the deep exposures.  For the typical cluster, 
we reach about 6 magnitudes below the turn-off, to about 0.2 $M_{\odot}$.
Table~\ref{tab01} provides the details of our observations for each cluster.

\begin{table}
\begin{center}
\caption{Summary of cluster observations.  All observations were taken in 2006. \medskip }
\label{tab01}
\begin{tabular}{|l|cc|rr|r|r|r|}
\hline 
\multicolumn{1}{|c|}{Cluster} &
\multicolumn{1}{ c }{Dataset} &
\multicolumn{1}{ c|}{Date} &
\multicolumn{1}{ c }{RA} &
\multicolumn{1}{ c|}{Dec} &
\multicolumn{1}{ c|}{PA\_V3} &
\multicolumn{1}{ c|}{F606W} &
\multicolumn{1}{ c|}{F814W} \\
\hline 
 Arp2    & j9l925 & 4/22 & 19:28:44 & $-$30:21:14 &  83.24 &  40s, 5x345s &  40s, 5x345s \\
 E3      & j9l906 & 4/15 & 09:20:59 & $-$77:16:57 & 245.09 &   5s, 4x100s &   5s, 4x100s \\ 
 Lynga7  & j9l904 & 4/07 & 16:11:02 & $-$55:18:52 & 124.32 &  35s, 5x360s &  35s, 5x360s \\ 
 NGC0104 & j9l960 & 3/13 & 00:24:05 & $-$72:04:51 & 346.17 &   3s, 4x 50s &   3s, 4x 50s \\ 
 NGC0362 & j9l930 & 6/02 & 01:03:14 & $-$70:50:54 &  44.24 &  10s, 4x150s &  10s, 4x170s \\ 
 NGC0288 & j9l9ad & 7/31 & 00:52:45 & $-$26:34:43 &  92.32 &  10s, 4x130s &  10s, 4x150s \\ 
 NGC1261 & j9l909 & 3/10 & 03:12:15 & $-$55:13:01 & 294.90 &  40s, 5x350s &  40s, 5x360s \\ 
 NGC1851 & j9l910 & 5/01 & 05:14:06 & $-$40:02:49 & 317.14 &  20s, 5x350s &  20s, 5x350s \\ 
 NGC2298 & j9l911 & 6/12 & 06:48:59 & $-$36:00:19 & 337.87 &  20s, 5x350s &  20s, 5x350s \\ 
 NGC2808 & j9l947 & 3/01 & 09:12:02 & $-$64:51:46 & 205.10 &  23s, 5x360s &  23s, 5x370s \\ 
 NGC3201 & j9l946 & 3/14 & 10:17:36 & $-$46:24:39 & 205.05 &   5s, 4x100s &   5s, 4x100s \\ 
 NGC4147 & j9l949 & 4/11 & 12:10:06 & $+$18:32:31 & 343.98 &  50s, 5x340s &  50s, 5x340s \\ 
 NGC4590 & j9l932 & 3/07 & 12:39:27 & $-$26:44:33 & 142.48 &  12s, 4x130s &  12s, 4x150s \\ 
 NGC4833 & j9l931 & 6/02 & 12:59:34 & $-$70:52:29 & 296.05 &  10s, 4x150s &  10s, 4x170s \\ 
 NGC5024 & j9l950 & 3/02 & 13:12:55 & $+$18:10:08 &  77.47 &  45s, 5x340s &  45s, 5x340s \\ 
 NGC5053 & j9l902 & 3/06 & 13:16:27 & $+$17:41:52 &  73.41 &  30s, 5x340s &  30s, 5x350s \\ 
 NGC5139 & j9l9a7 & 7/22 & 13:26:45 & $-$47:28:36 & 290.54 &   4s, 4x 80s &   4s, 4x 90s \\ 
 NGC5272 & j9l953 & 2/20 & 13:41:11 & $+$28:22:31 &  81.00 &  12s, 4x130s &  12s, 4x150s \\ 
 NGC5286 & j9l912 & 3/03 & 13:46:26 & $-$51:22:23 & 133.74 &  30s, 5x350s &  30s, 5x360s \\ 
 NGC5466 & j9l903 & 4/12 & 14:05:27 & $+$28:32:04 &  20.07 &  30s, 5x340s &  30s, 5x350s \\ 
 NGC5904 & j9l956 & 3/13 & 15:18:33 & $+$02:04:57 &  92.14 &   7s, 4x140s &   7s, 4x140s \\ 
 NGC5927 & j9l914 & 4/13 & 15:28:00 & $-$50:40:22 & 138.13 &  30s, 5x350s &  25s, 5x360s \\ 
 NGC5986 & j9l915 & 4/16 & 15:46:03 & $-$37:47:09 & 126.51 &  20s, 5x350s &  20s, 5x350s \\ 
 NGC6093 & j9l916 & 4/09 & 16:17:02 & $-$22:58:30 & 101.42 &  10s, 5x340s &  10s, 5x340s \\ 
 NGC6101 & j9l917 & 5/31 & 16:25:48 & $-$72:12:06 & 181.91 &  35s, 5x370s &  35s, 5x380s \\ 
 NGC6121 & j9l964 & 3/05 & 16:23:35 & $-$26:31:31 &  99.90 & 1.5s, 2x 25s,& 1.5s, 4x 30s \\ 
         &        &      &          &             &        &       2x 30s &              \\ 
 NGC6144 & j9l943 & 4/15 & 16:27:14 & $-$26:01:29 & 103.41 &  25s, 5x340s &  25s, 5x350s \\ 
 NGC6171 & j9l933 & 3/30 & 16:32:31 & $-$13:03:12 &  93.29 &  12s, 4x130s &  12s, 4x150s \\ 
 NGC6205 & j9l957 & 4/02 & 16:41:41 & $+$36:27:36 &  66.23 &   7s, 4x140s &   7s, 4x140s \\ 
 NGC6218 & j9l944 & 3/01 & 16:47:14 & $-$01:56:51 &  97.68 &   4s, 4x 90s &   4s, 4x 90s \\ 
 NGC6254 & j9l962 & 3/05 & 16:57:08 & $-$04:05:57 &  96.12 &   4s, 4x 90s &   4s, 4x 90s \\ 
 NGC6304 & j9l918 & 4/14 & 17:14:32 & $-$29:27:44 &  98.88 &  20s, 5x340s &  20s, 5x350s \\ 
\hline 
\end{tabular}
\end{center}
\end{table}

\begin{table}
\begin{center}
\medskip
\begin{tabular}{|l|ll|rr|r|r|r|}
\hline 
\multicolumn{1}{|l|}{Cluster} &
\multicolumn{1}{ c }{Dataset} &
\multicolumn{1}{ c|}{Date} &
\multicolumn{1}{ c }{RA} &
\multicolumn{1}{ c|}{Dec} &
\multicolumn{1}{ c|}{PA\_V3} &
\multicolumn{1}{ c|}{F606W}  &
\multicolumn{1}{ c|}{F814W} \\
\hline 
 NGC6341 & j9l958 & 4/11 & 17:17:07 & $+$43:08:11 &  62.25 &   7s, 4x140s &   7s, 4x150s \\ 
 NGC6352 & j9l959 & 4/10 & 17:25:29 & $-$48:25:22 & 105.79 &   7s, 4x140s &   7s, 4x150s \\ 
 NGC6362 & j9l934 & 5/30 & 17:31:54 & $-$67:02:53 & 106.79 &  10s, 4x130s &  10s, 4x150s \\ 
 NGC6366 & j9l907 & 3/30 & 17:27:44 & $-$05:04:36 &  87.53 &  10s, 4x140s &  10s, 4x140s \\ 
 NGC6388 & j9l919 & 4/06 & 17:36:17 & $-$44:44:06 & 100.71 &  40s, 5x340s &  40s, 5x350s \\ 
 NGC6397 & j9l965 & 5/29 & 17:40:41 & $-$53:40:24 & 148.54 &   1s, 4x 15s &   1s, 4x 15s \\ 
 NGC6441 & j9l951 & 5/28 & 17:50:12 & $-$37:03:04 & 122.48 &  45s, 5x340s &  45s, 5x350s \\ 
 NGC6496 & j9l9a9 & 5/31 & 17:59:03 & $-$44:15:58 & 134.74 &  30s, 5x340s &              \\ 
         & j9l920 & 4/01 & 17:59:03 & $-$44:15:58 &  94.46 &              &  30s, 5x350s \\ 
 NGC6535 & j9l935 & 3/30 & 18:03:50 & $-$00:17:48 &  86.04 &  12s, 4x130s &  12s, 4x150s \\ 
 NGC6541 & j9l936 & 4/01 & 18:08:02 & $-$43:42:57 &  92.60 &   8s, 4x140s &   8s, 4x150s \\ 
 NGC6584 & j9l921 & 5/27 & 18:18:37 & $-$52:12:54 & 131.18 &  25s, 5x350s &  25s, 5x360s \\ 
 NGC6624 & j9l922 & 4/14 & 18:23:40 & $-$30:21:39 &  90.06 &  15s, 5x350s &  15s, 5x350s \\ 
 NGC6637 & j9l937 & 5/22 & 18:31:23 & $-$32:20:53 &  99.28 &  18s, 5x340s &  18s, 5x340s \\ 
 NGC6652 & j9l938 & 5/27 & 18:35:45 & $-$32:59:24 & 101.82 &  18s, 5x340s &  18s, 5x340s \\ 
 NGC6656 & j9l948 & 4/01 & 18:36:24 & $-$23:54:12 &  86.47 &   3s, 4x 55s &   3s, 4x 65s \\ 
 NGC6681 & j9l939 & 5/20 & 18:43:12 & $-$32:17:30 &  96.43 &  10s, 4x140s &  10s, 4x150s \\ 
 NGC6715 & j9l923 & 5/25 & 18:55:03 & $-$30:28:41 &  94.18 & \multicolumn{1}{l|}{2x30s,} 
                                                           & \multicolumn{1}{l|}{2x30s,} \\ 
         &        &      &          &             &        &      10x340s &      10x350s \\ 
 NGC6717 & j9l940 & 3/29 & 18:55:06 & $-$22:42:03 &  84.61 &  10s, 4x130s &  10s, 4x150s \\ 
 NGC6723 & j9l941 & 6/02 & 18:59:33 & $-$36:37:54 & 106.02 &  10s, 4x140s &  10s, 4x150s \\ 
 NGC6752 & j9l966 & 6/24 & 19:10:52 & $-$59:59:04 & 119.42 &   2s, 4x 35s &   2s, 4x 40s \\ 
 NGC6779 & j9l905 & 5/11 & 19:16:35 & $+$30:11:05 &  59.22 &  20s, 5x340s &  20s, 5x350s \\ 
 NGC6809 & j9l963 & 4/19 & 19:39:59 & $-$30:57:44 &  81.46 &   4s, 4x 70s &   4s, 4x 80s \\ 
 NGC6838 & j9l9a8 & 5/12 & 19:53:46 & $+$18:46:42 &  65.46 &   4s, 4x 75s &   4s, 4x 80s \\ 
 NGC6934 & j9l927 & 3/31 & 20:34:11 & $+$07:24:15 &  89.61 &  45s, 5x340s &  45s, 5x340s \\ 
 NGC6981 & j9l942 & 5/17 & 20:53:27 & $-$12:32:12 &  72.88 &  10s, 4x130s &  10s, 4x150s \\ 
 NGC7078 & j9l954 & 5/02 & 21:29:58 & $+$12:10:01 &  77.40 &  15s, 4x130s &  15s, 4x150s \\ 
 NGC7089 & j9l952 & 4/16 & 21:33:26 & $-$00:49:23 &  78.04 &  20s, 5x340s &  20s, 5x340s \\ 
 NGC7099 & j9l955 & 5/02 & 21:40:22 & $-$23:10:45 &  69.63 &   7s, 4x140s &   7s, 4x140s \\ 
 Pal1    & j9l901 & 3/17 & 03:33:23 & $+$79:34:50 & 236.85 &  15s, 5x390s &  15s, 5x390s \\ 
 Pal2    & j9l908 & 8/08 & 04:46:06 & $+$31:22:51 &  87.56 &       5x380s &       5x380s \\ 
 Pal12   & j9l928 & 5/21 & 21:46:38 & $-$21:15:03 &  63.70 &  60s, 5x340s &  60s, 5x340s \\ 
 Terzan7 & j9l924 & 6/03 & 19:17:43 & $-$34:39:27 &  99.90 &  40s, 5x345s &  40s, 5x345s \\ 
 Terzan8 & j9l926 & 6/03 & 19:41:44 & $-$34:00:01 &  95.01 &  40s, 5x345s &  40s, 5x345s \\
\hline 
\end{tabular}
\end{center}
\end{table}

\begin{figure}
\plotone{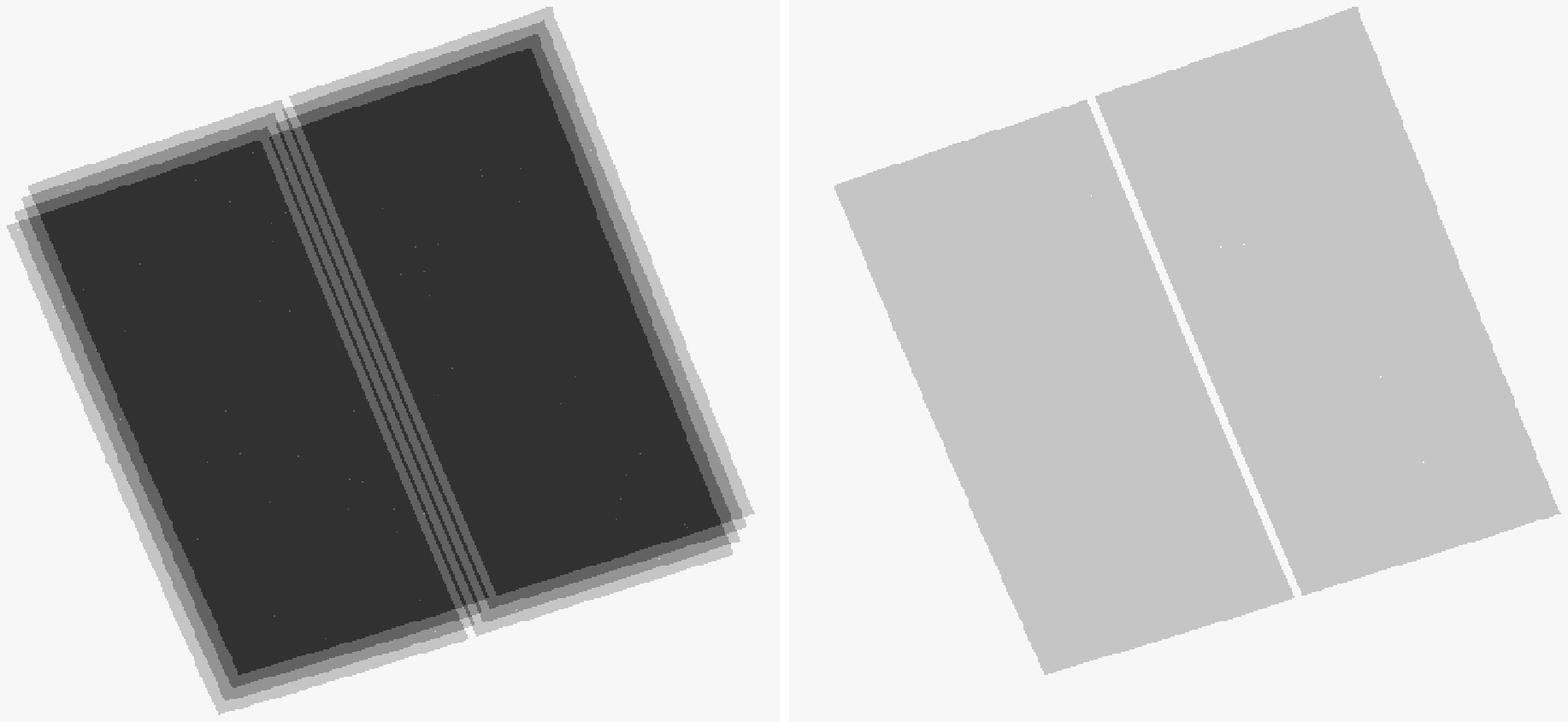}
\caption{{\it Left}: the depth of the deep stack in the case of four
         deep images.  Most parts of the field are covered by all four
         images.  We dithered the observations to ensure that a star
         will fall in the gap in at most one deep exposure; hence we
         have at least three images covering all areas but the very
         edges.  {\it Right}: the depth of the short stack (1 or 0).
         Some bright stars will fall in the gap of the short-exposure
         image and can be measured (albeit poorly) only in the deep
         exposures.
         \label{fig01}}
\end{figure}

To give the survey as much spatial uniformity as possible, we stepped
our observations so that no star would fall in the inter-chip gap in more 
than one of the deep exposures.  Since the WFC field-of-view is actually 
quite rhombus-shaped, we also made sideways steps so that the resulting 
field would be as square as possible.  Figure~\ref{fig01} shows the 
coverage for a typical cluster that had four deep exposures.

\section{PRELIMINARY SET-UP}
\label{s.SETUP}

The HST pipeline generates two main types of output image.  The {\tt
flt} images have been flat-fielded and bias-subtracted, but are
otherwise left in the raw WFC CCD frame, which suffers from a lot of
distortion.  The standard pipeline also generates a {\tt drz} image for
each set of associated exposures.  This is a drizzled, composite image
of all the exposures that were taken in the same visit through the same
filter.  The {\tt drz} images have been resampled into a standard
distortion-free frame and tied to an absolute astrometric frame via the
guide stars.  A careful photometric calibration has also been worked out
for them (Sirianni et al.\ 2005).  Thus, the {\tt drz} images can serve
to establish both our astrometric reference frame and photometric zero
points.  However, because they have been resampled, they are not well
suited for high-accuracy PSF-fitting analysis.  For this reason, we used
the {\tt drz} images for calibration, but our final measurements came
from careful analysis of the individual {\tt flt} images.

The first step in reducing the data for each cluster was to construct 
a reference frame and relate each {\tt flt} exposure to this frame, both 
astrometrically and photometrically.

\subsection{Constructing a reference frame for each cluster}
\label{ss.ref_frame}

To construct an astrometric frame for each cluster, we first measured 
simple centroid positions for the bright, isolated stars in the F606W 
{\tt drz} image.  Using the WCS header information, we converted these 
positions into a reference frame that has the targeted cluster center at 
coordinate [3000,3000], the $y$ axis aligned with North, and a scale of 
50 mas/pixel.  For all cluster orientations, this allows the entire observed 
field to fit conveniently within a frame that is 6000$\times$6000 pixels.

The next step was to relate each of the individual {\tt flt} exposures
to this reference frame.  We started by measuring positions and fluxes
for all of the reasonably bright stars in each {\tt flt} exposure with
the program {\tt img2xym\_WFC.09x10}, documented in Anderson \& King
(2006, AK06).  Briefly, the program starts with a library PSF, which was
constructed empirically for each filter using GO-10424 observations of
the outskirts of NGC~6397.  These library PSFs account for the spatial
variations in the WFC PSF due to the telescope optics and the variable
charge diffusion present in the CCD (see Krist 2005).  The PSF in each
exposure can differ from this library PSF due to spacecraft breathing or
focus changes, so we fitted the library PSF to bright stars in each
image and came up with a spatially constant perturbation to the PSF that
better represents the star images in each individual exposure.  Using
the improved PSF, the program then went through each exposure and
measured positions and fluxes for the bright, isolated stars.  The
exposure-specific, improved PSFs were saved for later in the analysis.

We next found the common stars between the reference list and the star list 
for each exposure.  This allowed us to define a general, 6-parameter linear 
coordinate transformation from the distortion-corrected frame of each 
exposure into the reference frame.  Since the photometry and astrometry are
more accurately measured in the {\tt flt} frames than in the {\tt drz} 
frame, we improved the internal quality of the reference frame by 
iteration.  The final reference-frame positions and fluxes for the bright 
stars should be internally accurate to better than 0.01 pixel and 
0.01 magnitude.

Using this reference list of stars for each cluster, we computed
the final astrometric transformations and photometric zero points from 
each short and deep exposure into the reference frame.  The photometric 
system at this stage was kept in instrumental magnitudes, 
$-2.5\,{\rm log}_{10}({\rm flux}_{\rm DN})$, where the flux corresponds 
to that measured in the deep {\tt flt} images for the cluster at hand.
It was convenient to keep our photometry in this instrumental system until 
calibration at the very end (\S~\ref{ss.photo_calib}), because instrumental 
magnitudes make it easier to assess errors in terms of the expected 
signal to noise.

\subsection{Stack construction}
\label{ss.stacks}

The transformations from the individual exposures into the reference 
frame allowed us to construct a stacked representation of each field.  
We did not use these stacks in the quantitative analysis, but they were 
an invaluable tool which enabled us to inspect star lists and evaluate
the star-finding algorithm.  (It is worth noting that the {\tt drz} 
images which were produced in the ACS pipeline were not adequate for 
this for several reasons:  
[1] the pipeline uses the commanded POS-TARGs to register the exposures in
    a common frame, whereas our empirical star-based transformations allow
    a much more accurate mapping from the exposures into the reference frame; 
[2] the pipeline is set up to deal with an arbitrary set of images with
    different exposure times, whereas our stacking algorithm could be 
    optimized for the 3 to 5 deep exposures plus one shallow exposure 
    that we have for each filter; and 
[3] we wanted the image to be in our reference frame, but did not want 
    to resample the {\tt drz} image and thus degrade the resolution even 
    further.)

There is no unique way to construct a stacked image from a dithered set 
of exposures.  Our construction of the stacks was analogous to using 
{\tt drizzle} (Fruchter \& Hook 2002) with {\tt pixfrac} = 0.  We went
through the reference frame pixel by pixel and used the inverse 
coordinate transformations and inverse distortion corrections to map the 
center of each reference-frame pixel into the frame of each of the 
individual F606W exposures.  We then identified the closest pixel in each 
of the 3 to 5 exposures and computed a sigma-clipped mean of these pixel
values.
Finally, we set the value of the reference-frame pixel to this mean, and
moved on to the next pixel in the reference frame.  This produced a
stack of the deep exposures.

\begin{figure}
\plotone{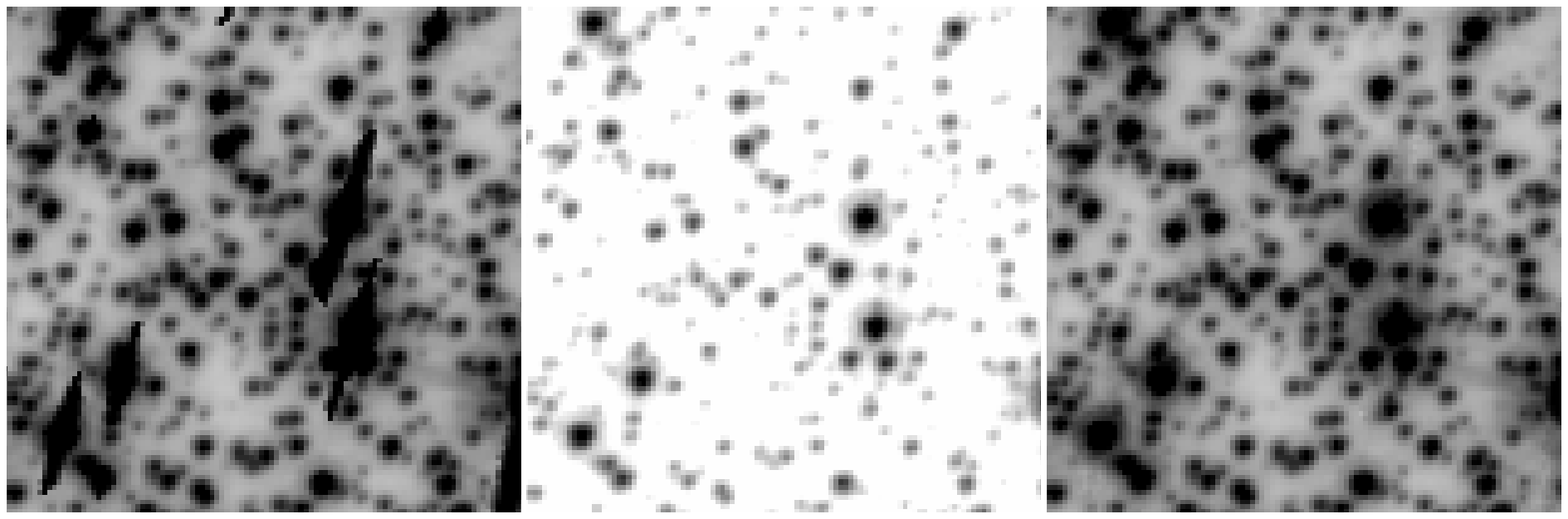}
         \caption{({\it Left}) A 100$\times$100-pixel (5$\times$5-arcsecond) 
                               region in the stack from the deep F606W 
                               exposures of 47 Tuc;
                ({\it middle}) same for the short F606W exposure;  
                 ({\it right}) the combination of the long and short exposures.
         \label{fig02}}
\end{figure}

To deal with pixels that were saturated in the deep exposures, we 
generated a similar stack from the short exposure (actually, a stack 
from just one image is better called a resampling).  We then constructed a 
composite stack by starting with the deep-exposure stack and replacing 
any pixel that was within 3 pixels of a saturated pixel with the 
exposure-time-scaled value from the short-exposure stack.  Finally, 
we put the WCS header information into this composite stack for each 
filter.  Figure~\ref{fig02} shows an example of the stacked images for 
a 100$\times$100-pixel region at the center of 47 Tuc.  

We constructed such a composite stack for the F606W and the F814W 
exposures for each cluster.  These stacks were not used directly in 
the reductions discussed in the next sections, but because they are a 
simple representation of the scene without regard to the locations of 
stars, they provide a critical sanity test of our finding and measuring 
routines.  They will also serve as excellent finding charts for future 
spectroscopic projects.


\section{GENERAL CONSIDERATIONS FOR THE FINDING AND MEASURING PROCEDURE}
\label{s.OVERVIEW}

In the previous section, we constructed a calibrated reference frame 
for each cluster and found the photometric and astrometric transformation 
from each exposure into this frame.  These transformations allowed us 
to construct a composite stacked image for each cluster.  The next step 
was to construct a composite list of stars for each cluster.

Our strategy for finding and measuring stars had to be tailored to the 
scientific goals of the project and to the specifics of the detector and 
fields.  In this section, we discuss some of the issues involved in 
constructing a catalog of stars from moderately undersampled images of 
globular clusters, where the stellar density can vary by orders of 
magnitude, and where there are both bright giants and faint main-sequence 
stars together in the same field.  In this section, we provide an overview 
of the reduction; the details will be given in Section \ref{s.KSYNC}.  

\subsection{The goals of the survey}
\label{ss.GOALS}
There are many different scientific objectives for this data set:  
luminosity-function analysis, isochrone fitting, binary studies, etc.  
Many of these different applications would benefit from different sampling 
strategies.  For instance, luminosity-function (LF) studies do not require 
precise photometry to sift stars into 0.5-magnitude-wide bins, but LF 
studies {\it do} depend on high completenesses and reliable completeness 
corrections.  On the other hand, when fitting isochrones to CMDs, we do 
not need a particularly complete sample of stars, but we do need a sample 
with the smallest possible photometric errors.  In order to satisfy these 
competing requirements, we pursued a two-pronged strategy. Our primary goal was 
to identify as many stars as possible, so that no future searches would be 
necessary on these images.  At the same time we sought to document which 
stars were more likely to be better measured.  This way, each application can 
cull from the catalog the sample of stars that is best suited for the 
analysis at hand.

\subsection{The need for automation}
\label{ss.AUTOMATION}
While we wanted our catalogs to be as comprehensive as possible, because 
of crowding and signal-to-noise limitations we could not hope to identify
every star in every cluster field.  The best we could hope for was to find 
all the stars that could be found by a careful human.  
There are hundreds of thousands of stars in many of these fields, so 
finding stars by hand was not very practical.  Add to this the need 
to run artificial-star tests and it was clear that we had to come up 
with a completely automated finding and measuring procedure.  This 
procedure had to:  
(1) be optimized for the WFC detector and globular-cluster fields, 
(2) find almost everything that a person would find, 
(3) misidentify a minimum of artifacts as stars, and 
(4) measure each star as accurately as possible.  Below we discuss our
general approach to dealing with these issues.  In the next section, we
will deal with the specifics.

\subsection{Finding stars in undersampled images}
\label{ss.finding_undersam}
In well-sampled images, it can be useful to convolve the image with a PSF 
in order to highlight the signal from the point sources over the random 
pixel-to-pixel noise.  In undersampled images, however, much of the flux 
of a point source is concentrated in its central pixel.  This undersampling 
makes it counterproductive to convolve the image before finding, because 
the stars already stand out as starkly as they can in the raw frames 
(or in frames in which the brighter stars have been subtracted out).  
An additional complication of undersampling 
is that it is often difficult to determine from a single undersampled 
image whether or not a given detection is stellar, so we need some
independent way to establish which detections are really stars.  

The best way to find stars in undersampled images, then, is to take 
a set of dithered exposures and look for significant local maxima 
(or ``peaks'') that occur in the same place in the field in several 
independent exposures.  The dithering is critical because it allows 
us to differentiate real sources from warm pixels or cosmic rays.

\subsection{Iterative finding}
\label{ss.finding_human}
The stars used in \S\ \ref{ss.ref_frame} to relate the individual exposures 
to the reference frame were found with a single pass through each image.
The finding routine found only stars that had no brighter neighbors 
within four pixels.  Such an algorithm finds almost all of the bright
stars in a field, but it misses many of the obvious faint stars in the 
wings of the bright ones.  If after finding the bright stars, we were then 
to subtract them out and search for more stars in the subtracted images, 
we could both find more faint stars and at the same time improve the 
photometry for the brighter stars (by subtracting the fainter stars before
our final measurement of the brighter ones).

There are two ways to perform such an iterative search.  The first approach 
is to make multiple passes through the entire field.  This has the advantage 
that it treats the field as a contiguous unit, but it is extremely memory 
intensive and requires maintaining many large, intermediate images (the raw 
images, subtracted images, model images, etc). 

An alternative strategy is to reduce one patch of the field at a time, doing 
multiple passes on that patch before moving on to the next patch.   Such a
patch must be larger than the distance over which stars can influence 
each other, but it can be small enough to allow the transformations 
to be linear and to treat the PSFs as spatially constant within the patch.  
The patch approach also has an advantage for doing artificial-star (AS)
tests.  When reducing the entire field as a unit, AS tests must be done 
in parallel.  To ensure that artificial-stars will not affect the crowding 
they are intended to measure, we can add at most one test star 
every 20$\times$20 pixels and are thus limited to about $\sim$40,000 stars 
per run.  On the other hand, with a patch-based approach we can do 
AS tests in series, one after another, with no worry of them ever 
interfering with each other.  This allows the number of tests per run 
to be limited only by computing time.  For all these reasons, we 
chose to reduce each field using a mosaic of local patches.
The details of this will be fleshed out in \S~\ref{ss.patch}.

\subsection{Avoiding artifacts}
\label{ss.avoiding_artifacts}
One of the complications of studying globular clusters is that there are 
almost always very bright stars and very faint stars in the same field,
and we want to study them both.  The bright stars affect the faint stars
in two ways.  First, they dominate the region closest to them, making it 
hard to find faint stars that are too close.  But the extremely bright 
stars also affect an even larger region around them because of the mottled 
wings of the PSF, which are very hard to model accurately.

\begin{figure}
\plotone{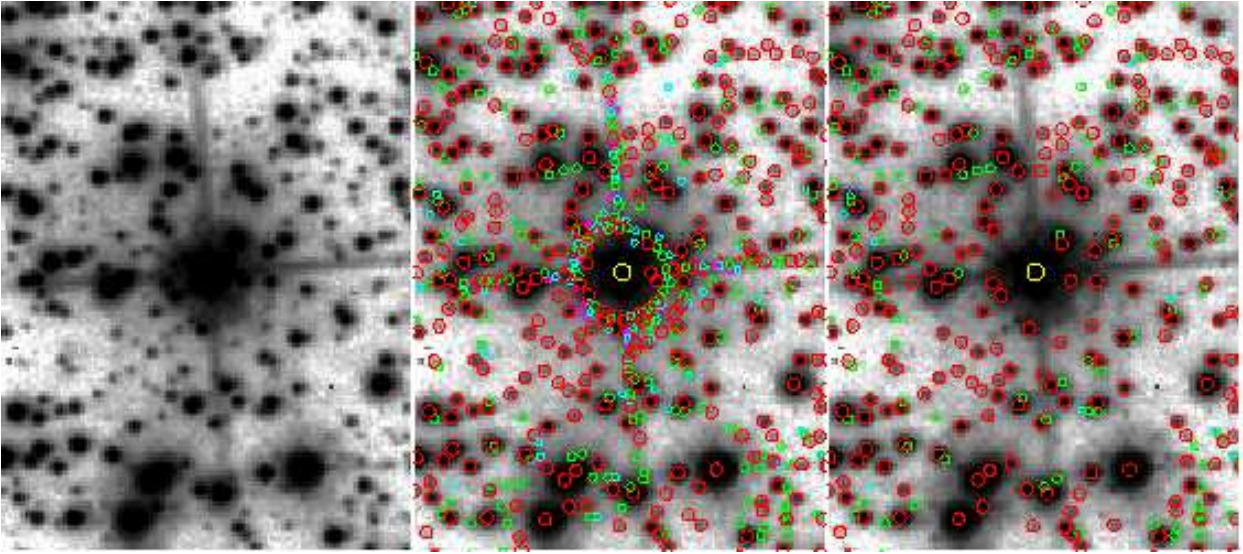}
\caption{The stars found in a 120$\times$160-pixel region of NGC~6715 in
         the vicinity of a bright star.  The left panel shows the scene 
         without any stars indicated.  The middle panel shows all the 
         ``stars'' that would be found if we did not consider the influence 
         of the bright star.  The right panel shows the stars that were 
         found to be bright enough to be distinct from the profile of the
         bright star (see \S\ \ref{ss.setting_up}).  The different colors 
         and sizes of the circles correspond to the different passes through 
         the data.  The large yellow circles are stars that were found in 
         the first pass. These stars are saturated in the deep exposures.  
         The red symbols were found in the first deep-exposure pass, with 
         the increasingly smaller green, cyan, and magenta symbols indicating 
         stars found in subsequent passes.  
         \label{fig03}} 
\end{figure}

To ensure that false detections, such as PSF artifacts or residuals from 
imperfect subtraction of bright stars, would not enter into our sample, we 
ended up insisting that any new stellar detection must stand above a 
conservative estimate of the error in the subtraction of the previously 
identified brighter stars.  In practice, this means that there is a limit 
to how close to a brighter star a given fainter star can be reliably 
found.  We determined that while such a requirement does exclude a small number 
of stars that could have marginally been found by hand, it does an 
excellent job of excluding non-stellar artifacts from the sample
(see Fig.~\ref{fig03} and \S~\ref{ss.setting_up}).  The region of 
exclusion as a function of brightness can easily be quantified by 
artificial-star tests.

\subsection{Measuring stars in undersampled images}
\label{ss.measuring_undersam}
Once stars were found, we had to measure fluxes and positions for them,
and our measuring algorithms also had to be tailored to the particulars 
of the detector and fields.  Most of the signal from stars in undersampled 
images is concentrated in the stars' central few pixels, so our fits 
clearly had to focus on those pixels.

There are two ways to measure stars in multiple exposures.  We can 
either measure each star independently in each exposure and later combine
these observations, or we can fit for a single flux and position for each 
star simultaneously to all the pixels in all the exposures.  The first 
approach is generally better for bright stars, where each exposure presents 
a well-posed problem with an obvious stellar profile to fit.  The latter 
approach is better for very faint stars, which cannot always be 
robustly found and measured in every individual exposure.  In our 
procedures, we ended up computing the flux for each star both ways.  
The vast majority of stars we found were bright enough to be measured 
well using the first approach, so our basic catalog reports just the 
independently fitted fluxes.  We did, however, save the simultaneous-fitted 
fluxes in auxiliary files.

It is worth noting that although our aim was to construct a uniform sample, 
it was not possible to measure all stars with the same quality.  Some stars 
were bright and isolated and could be measured with a large, generous 
fitting radius.  Other stars were crowded or faint, and only 
their core pixels could be fitted.  In a sense, each star presented a special 
circumstance, and our general measuring algorithm had to be able to adapt 
as much as possible to minimize the most relevant errors for each star.  
The PSF provided the unifying measuring stick that enabled us to evaluate 
a consistent flux for all the stars, even though the fit to different stars 
sometimes had to focus on different pixels.

\subsection{Summary of the considerations}
\label{ss.consid_summy}
In summary, our finding and measuring strategy had to take into account 
the nature of the data set and the goals of the survey.  We clearly needed
an automated procedure that could find stars simultaneously in multiple
dithered exposures.  The procedure would have to be able to use multiple 
iterative passes to identify faint stars in the midst of brighter ones, 
and it would also have to be robust against inclusion of PSF artifacts or 
subtraction residuals as stars.  Finally, we needed to come up with a way 
to measure a flux and position for each star, taking into consideration 
its particular local environment.


\section{THE REDUCTION PROGRAM}
\label{s.KSYNC}
We designed a sophisticated computer program ({\tt multi\_phot\_WFC}) 
that could deal with all of the above requirements in a generalized way, 
so that the same program could be used to reduce the data for every 
cluster in the sample, no matter how much crowding or saturation the 
cluster might suffer at its center.  The program takes as input the 
10 to 12 raw {\tt flt} images in each cluster's data set and the 
background information about how each exposure is related to the 
reference frame.  It then analyzes the images simultaneously and 
outputs a list of stars that it found, including a position, $V$ and $I$ 
photometry, and some data-quality parameters for each star.  It 
was set up to run in two different modes: finding real stars and 
running artificial-star (AS) tests.  In this section we describe the 
mechanics of the real-star search.  In \S\ \ref{s.ASTEST} we discuss 
the AS operation, which differs only in the set-up and the output stages.

\subsection{The patch}
\label{ss.patch}
We chose to reduce each cluster field one patch at a time, both to
conserve memory and to facilitate artificial-star tests.  The size of
the patch was a compromise between the desire to cover as much field as
possible in each patch in the real-star runs without covering too much
unnecessary field in the artificial-star runs.  We thus arrived at a
patch size of 25$\times$25 pixels.  Since stars at the edge of a patch
often have significant neighbors outside of the patch, each patch
allowed us to fully treat only its central 11$\times$11-pixel region.
We centered a patch every 10$\times$10 pixels, so the entire
6000$\times$6000-pixel reference frame for each cluster was covered by an 
array of 600$\times$600 patches.

To set up each patch for analysis, we used the transformations from
\S\ \ref{ss.ref_frame} to map the location of the central pixel of the patch 
into each of the exposures, and extracted a local 25$\times$25-pixel raster from 
each exposure.  We constructed a PSF model for each exposure using the 
appropriate library PSF for that location on the chip and the perturbation 
component found in \S\ \ref{ss.ref_frame}.  

We also determined the linear transformation from the patch frame into
the raster for each exposure.  Using these transformations, we
intercompared the pixels for the individual $F606W$ and $F814W$ rasters
and flagged as bad any pixels that were discordant by more than
$5\sigma$ with the other images for that filter.  We inspected a large
sample of the resulting rasters and verified that the obvious cosmic
rays and warm pixels were identified.  This procedure enabled us to do
simultaneous fits to all the exposures without having to check each time
for bad pixels.

\subsection{Setting up the bright-star mask}
\label{ss.setting_up}

One of the challenges in constructing a catalog from this 
data set was to avoid including PSF artifacts as stars.  
Stars brighter than an instrumental magnitude of $-12.5$  
($10^5$ $e^-$ total) often have knots and ridges in their 
PSFs that can be confused with stars.  These features are 
hard to model accurately, and therefore cannot be  
subtracted off well.  The best we could do was to 
conservatively estimate their contaminating influence 
and make sure that the stars we found stood out clearly 
above the bright-star halos.

To do this, we identified several bright, isolated stars that
were highly saturated in the deep exposures and examined
their radial profiles.  Since we had a flux for each bright star
from the short exposures, we could examine the radial profile
for the star with a scaling matched to the PSF.  Ignoring 
for now the diffraction spikes, we looked at the envelope of the
trend with radius and drew by eye a curve that encompassed all 
of the obvious halo structure.  Since the halo structure is 
largely due to scattered light, it should have the same level 
in F606W and F814W, though the detailed structure will be 
different for the different filters.  
Table~\ref{tab02} gives the upper-envelope profile we found
in the $f_{\rm any}$ column.

\begin{table}
\begin{center}
\caption{The profile in the vicinity of a bright star below which 
         a fainter star is likely to be confused with a PSF artifact.  
         Outside of the core, this is essentially a generous upper limit 
         for the PSF as a function of radius.  The PSF is normalized
         to have a flux of 1.0 within 10 pixels.   The ``any'' column
         refers to the general radial trend, while the ``spike'' column 
         tracks how the PSF intensity varies with radius along the 
         spikes. \bigskip }
\label{tab02}
\begin{tabular}{|c|rr||c|rr|}
\hline 
Radius & $f_{\rm any}$ {\ \ }   & $f_{\rm spike}$ {\ \ }  &
Radius & $f_{\rm any}$ {\ \ }   & $f_{\rm spike}$ {\ \ }  \\
\hline
  0 & 0.005000 & 0.005000 & 55 & 0.000000 & 0.000015 \\
  5 & 0.002500 & 0.002500 & 60 & 0.000000 & 0.000010 \\
 10 & 0.000300 & 0.000300 & 65 & 0.000000 & 0.000008 \\
 15 & 0.000075 & 0.000100 & 70 & 0.000000 & 0.000006 \\
 20 & 0.000035 & 0.000050 & 75 & 0.000000 & 0.000005 \\
 25 & 0.000020 & 0.000042 & 80 & 0.000000 & 0.000004 \\
 30 & 0.000010 & 0.000035 & 85 & 0.000000 & 0.000003 \\
 35 & 0.000005 & 0.000030 & 90 & 0.000000 & 0.000002 \\
 40 & 0.000003 & 0.000025 & 95 & 0.000000 & 0.000001 \\
 45 & 0.000002 & 0.000022 &100 & 0.000000 & 0.000000 \\
 50 & 0.000001 & 0.000020 &    &          &          \\  
\hline
\end{tabular}
\end{center}
\end{table}

To make use of this profile,
before we began the finding procedure for each patch, we
first identified all the bright stars that might generate 
artifacts that could be confused with stars in the patch 
by determining which stars in the bright reference list
(\S~\ref{ss.ref_frame}) were within 100 pixels of the patch.
For each of these nearby bright stars, we used Table~\ref{tab02}
to evaluate this upper-limit estimate for each pixel in the 
raster for each exposure, based on the radial distance and 
total flux of the bright star.  This was recorded in a 
separate raster called the ``mask'' raster.  Later, when we 
searched for stars, we required that a star stand out above this 
level to be considered a possible stellar detection.

The above treatment did not address the diffraction spikes.
Without masking them out also, an automated, multi-pass
routine would tend to find beads of false stars along the
spikes.  The spikes are complicated to deal with since that
they emanate from the bright stars at different angles with 
respect to the undistorted {\tt flt} pixel grid at different 
locations in the field (due to the large distortion in the 
WFC camera).  Since the changes in angle were small, the spikes 
were still largely directed along $x$ and $y$, at least over 
the short distance of a patch.  So when there was an extremely 
bright star within 100 pixels directly to the left or right or 
directly above or below the current patch, we looked for a linear 
ridge in the patch that was directed towards the bright star.  
Once the exact location of the spike was identified, we used the 
$f_{\rm spike}$ column of Table~\ref{tab02} to mask out the
relevant pixels.   The entries in this column were also 
constructed by examining the radial profiles of spikes around 
bright stars.

(We note that while the above approach successfully prevented
diffraction spikes from being identified as stars, we were
dissatisfied with the somewhat imprecise treatment of the spikes.
So in the time since the GO-10775 reduction, we have done a more
thorough characterization of the spikes' 
angles with chip location.  We verified that the spikes
are fixed relative to the detector and that their orientation
changes linearly with location on the chip such that, for
example, the spike along $x$ will be directed at $-4.0^{\circ}$ at the
upper-left corner of the 4096$\times$4096 detector, and at $-2.3^{\circ}$
at the upper-right corner.  We have now folded this more precise spike 
treatment into the reduction routine, so that when it is used on 
future data sets, the spike treatment will be more rigorous.
We reiterate, though, that the star lists presented here should be 
free of spike contamination.)

An additional step in the set-up was to deal with saturated stars.  Our
routines were designed to find stars by looking for local maxima in images.  
Saturated stars tend to have a plateau of saturated pixels at their 
centers, so they cannot be automatically identified as detections by 
peak-based algorithms.  So, in a pre-processing stage for each exposure, 
we examined each contiguous region of saturation and artificially added 
a peak at the center, so that the automated routine would know to find 
a star there.  The routine then fit the wings of the PSF to the unsaturated 
pixels, allowing us to include the bright stars in the star lists and 
luminosity functions.  While this wing-fitting approach is the only 
way to measure {\it positions} for saturated stars, we show in 
\S~\ref{ss.satphot} that there is a better way to measure accurate fluxes.

Finally, in addition to correcting the centers of saturated stars, we 
also made a ``saturation map,'' which showed how many of the deep images 
were saturated at each location within the patch.  The saturation could 
be either because of direct illumination or because of charge blooming.  
The saturation map helped us to know where in the patch we should trust 
the short exposures more than the deep ones.

\subsection{Finding stars in the patch}
\label{ss.finding_stars}
Once the rasters, transformations, PSFs, and other background information 
had been assembled for each patch, we were finally ready to find the stars.  
As we mentioned in \S\ \ref{ss.finding_human}, our aim was to construct as 
comprehensive a catalog as possible, so we could not afford to find just 
the ``easy'' stars, but rather we needed to find all the stars that could 
be reliably found.  Thus the finding process would have to involve 
multiple iterative passes in which we first found the brightest stars, 
subtracted them, then searched for additional stars in the residuals.  
The goal during this finding process was not to measure the most accurate 
flux and position possible.  At this stage, we simply needed a good basic 
idea of where all the stars were in the patch, and roughly how bright they 
were, so that when we later made our final measurements, we could measure 
each star better by removing a good model for the contribution of its 
neighbors.

At the beginning of each finding iteration, we constructed a model 
for the raster for each exposure, 
using the current list of stars and the appropriate PSF.  We subtracted 
this model from the original raster and also subtracted a sky value as 
determined from the entire raster.  This was the ``residual'' raster, and 
there was one for each exposure.  (The residual rasters for the first 
iteration were just the sky-subtracted raw images, since there were not
yet any sources to subtract.)

In each iteration, we constructed a map of potential new sources in the
patch by going through the residual raster for each exposure, pixel by
pixel.  If we found a peak that had: (1) at least 10$\times$ the sky
sigma in its brightest 2$\times$2 pixels, (2) no unsubtracted brighter
neighbors or saturated or bad pixels within 3.5 pixels, (3) at least
25\% more flux in its brightest 2$\times$2 pixels than the model of the
previously found stars predicted, and (4) more flux than the bright-star
mask at that point, then we considered it a possible stellar detection.
We added a `1' to the new-source map at the appropriate location in the
patch.  We also kept track of the particularly high-quality detections
(those that had a distinctively PSF shape) in a separate
high-quality-source map.  

Once we had gone through all the exposures, we scanned the new-source
map to see where in the patch multiple exposures might have detected the same stars.  For 
parts of the field where we had all 10 deep exposures available, a star had 
to be detected independently in at least 5 of them to qualify for the list.  
At the edges of the field, where we had coverage from only one or two 
F606W and F814W exposures, we could not rely on an abundance of
coincident detections to validate each star.  Yet we still wanted to find
the obvious stars in the outer regions.  So, we allowed for a lower 
threshold number of detections but insisted that the detections be 
``high-quality'' (having a good fit to the PSF).  Table~\ref{tab03} gives 
the number of detections required as a function of how many images 
were available.

\begin{table}
\begin{center}
\caption{Number of regular and high-quality individual-exposure 
         detections that are required to constitute a formal stellar detection 
         as a function of $N_{\rm deep}$, the number of deep images that 
         cover a given point in the field.  (A detection must satisfy one or 
         the other.) \medskip 
         \label{tab03} }
\begin{tabular}{|c|rr|}
\hline 
\multicolumn{1}{|c|}{$N_{\rm deep}$} & \multicolumn{1}{ c }{$N_{\rm
peak} $} & \multicolumn{1}{ c|}{$N_{\rm qual}$} \\ \hline 1 & --- & ---
\\ 2 & --- & 2 \\ 3 & 3 & 2 \\ 4 & 3 & 2 \\ 5 & 4 & 3 \\ 6 & 4 & 3 \\ 7
& 4 & 3 \\ 8 & 5 & 3 \\ 9 & 5 & 4 \\ 10 & 5 & 4 \\ \hline
\end{tabular}
\end{center}
\end{table}

After identifying a star in the patch, we then measured it by fitting the 
PSF to the star simultaneously in all the exposures, solving for four 
parameters:  an $x$ and a $y$ position, and a flux in each of F606W and F814W.
Identifying a new star next to a previously found star can affect the old 
star's flux, so we iterated the fitting process until we converged on a 
position and a $V$ and $I$ flux for each of the stars in the current list.
These simultaneous-fit positions are not the best way to measure all 
stars, but they do give us a robust starting point.  The iteration was 
completed when we had converged on flux and position estimates for all 
the currently known stars in the patch.

\subsection{The multiple passes through the patch}
\label{sss.multiple_passes}
The above narrative describes what happened each time we passed through
the patch looking for new stars.  During the first two passes, we 
searched only the short exposures, looking exclusively at the parts of 
the patch that were saturated in the deep exposures.   Thanks to the 
pre-processing of the saturated regions (see \S~\ref{ss.setting_up}) 
the automated procedure was able to identify saturated stars as well 
as unsaturated stars.  In the third pass, we looked at parts of the 
patch that were saturated in the deep exposures, but which had no 
short-exposure coverage (for example, if the patch happened to fall 
in the inter-chip gap of the short exposures, see Fig.~\ref{fig01}).
This way, we did not miss any of the brightest stars.  Finally, in the 
fourth and subsequent passes, we focused on unsaturated stars in the 
deep exposures.  We performed up to ten additional passes through the 
patch.  Usually, all of the stars were found after very few passes, but 
sometimes there were particularly crowded regions that required 
up to ten passes.  Once no additional stars were found at the end of 
a pass, we moved on to the measurement stage.

Figure~\ref{fig03} shows an example of the stars found in a region of
NGC~6715.  In the left panel, we show all the sources that would be found 
by our algorithm if we were to find everything that generated a significant 
number of peaks, without regard to the bright-star mask.  On the right, 
we show how well our bright-star mask rejected the non-stellar artifacts 
around bright stars.  The multiple-pass approach typically found two to 
three times more stars than the single-pass procedure that was used 
to identify the bright isolated stars in \S~\ref{ss.ref_frame}.  

\subsection{The measurement stage}
\label{ss.measurement}

Once we had a final list of stars, we sought to measure each one as
accurately as possible.  The simultaneous-fitting method used above works 
best for very faint stars (see \S~\ref{ss.measuring_undersam}), but 
the vast majority of stars in our catalog were bright enough to be found 
and measured well in the individual exposures.  So, we measured each star 
in each of the individual exposures where it could be found, after 
first subtracting off its neighbors.  We measured a sky value from 
an annulus between 3 and 7 pixels for the fainter stars 
and between 4 and 8 pixels for the brighter stars (an estimate of the star's 
own contribution is subtracted before the sky is measured).  We then fit 
the PSF to the star's central 5$\times$5 pixels, in the manner of AK06.  
This worked well for isolated stars, but if the known neighbors contributed 
more than 2.5\% of the flux in the 25-pixel aperture, we found it was better 
to concentrate the fit on the centermost pixels.  Such a weighted fit is 
more susceptible to errors in the PSF, but it is less susceptible to 
errors in modeling of the neighbors.

In this way, we obtained between three and five independent estimates for 
each star's position and flux in each of the two filters.  From these 
multiple estimates we computed an average position and flux and an 
empirical estimate of the errors.  We also constructed a few diagnostics 
related to the quality of each measurement.  We recorded $o_V$ and $o_I$, 
the fraction of flux in the aperture coming from known neighbors, and 
$q_V$ and $q_I$, which are derived from the fractional residuals in
the fit of the PSF to the pixels.  The first pair of parameters can help 
to select a subset of stars that are more isolated from nearby 
contaminating neighbors, and hence presumably better measured.  The 
second pair can also help to select isolated stars, but this time by 
highlighting the stars that are not parts of barely resolved, but not-easily-separable blends.  
Section~\ref{s.PHOTO_ERRORS} will illustrate some ways to use these quality 
parameters.

\subsection{Generating the output catalog}
\label{ss.list}
Only the stars in the central region of each patch could be optimally 
measured, since stars at the edges could have unaccounted-for neighbors 
just outside the patch.  So we added to our final list of sources all 
the stars within the central 11$\times$11 pixels of the 25$\times$25-pixel 
patch.  We used the transformations from \S~\ref{ss.patch} to convert 
the local positions into the reference frame.  To ensure that no star 
near the border of the patch would be counted twice, we only added stars 
to the final list that were not already in the list.

For each star, the main output file records:\  (1) a position in the 
reference frame, (2) an instrumental F606W and F814W magnitude, (3) errors 
in the positions and fluxes, (4) the number of images where the star could 
have been found, and the number in which it was actually found, (5) an 
estimate of the flux in the aperture coming from other stars, (6) an 
estimate of the quality of the PSF fit, and (7) the simultaneous-fit 
fluxes.  We also record how the star was found: whether (best-case 
scenario) it was found unsaturated in the multiple deep exposures, 
unsaturated in the short exposure, saturated in the short exposure, or 
(the worst-case scenario) it could only be found as saturated in the 
deep exposures, because it fell in the gap of the short exposure. 

Note that for each star we kept track of both the average fluxes from the 
individual-exposure measurements and the fluxes obtained from the 
simultaneous fitting to all exposures at once.  Since the vast majority of 
stars were bright enough to be measured well in each individual exposure, 
in the main catalog we report only the average fluxes.  But we do
record the simultaneous-fit fluxes in auxiliary files.  In addition, we
also preserve in auxiliary files the photometry from the individual 
exposures so that variable stars can be identified and studied.


\section{ARTIFICIAL-STAR TESTS}
\label{s.ASTEST}

The patch-based approach made it very easy to perform artificial-star tests.
The standard way of performing AS tests is to do them {\it in parallel}:  
several sets of images are doped with an array of artificial stars, which 
are far enough apart not to interfere with each other; the images are then 
reduced blindly and the output lists are matched against the input lists, 
to see which stars were found.  The patch-based approach allows us to do 
AS tests {\it in serial}, one artificial star at a time.  This allows us 
to do the whole set of AS tests in completely automatic fashion, and requires 
no auxiliary image files.

\subsection{One artificial star at a time}
\label{ss.one_at_a_time}
An artificial-star test asks the question: If a star of a particular
magnitude and color is added at a particular location in the field, will
it be found, and if so, what will its measured magnitude and color be?
To answer this question, we simply define a patch that is centered at
the target location (as in \S\ \ref{ss.patch}) and then add the star, 
with the appropriate scaling, PSF, and noise, into the raster for each 
exposure.  The patch is then reduced in a completely automatic way using 
the procedures described in the preceding section; this generates a list 
of all sources that were found and measured.  The AS routine then reports 
the star that was found closest to the inserted position.  Once this has 
been completed, the procedure can be repeated for the next artificial star.
These artificial stars can never interfere with each other, because each 
one is added only to the rasters, which are temporary copies of the exposures.

Each artificial-star test thus consists of a set of input parameters  
({\tt x\_in}, {\tt y\_in}, {\tt mv\_in}, and {\tt mi\_in}), and the same 
output parameters as in \S~\ref{ss.list} for the nearest found star.  The 
end user will later have to determine whether the recovered star corresponds 
to the inserted star.  Typically, if the input and output positions agree to 
within 0.5 pixel and the fluxes agree to within 0.75 magnitude, then the 
star can be considered found.  If the star was recovered much brighter, 
then that means it was inserted on top of a brighter star and was not found 
as itself.  Also, if it was recovered more than 0.5 pixel away, then it is 
likely that the star itself was not found, but a brighter nearby neighbor 
was.  It is of course equally necessary to deal with such issues 
in the ``parallel'' way of doing AS tests.

\subsection{The standard run of tests}
\label{ss.standard_run}
We generated a standard set of artificial-star tests for each cluster 
in order to probe our finding efficiency and measurement quality from 
the center to the edge of
the field.  We inserted the artificial stars with a flat luminosity 
function in F606W, with instrumental magnitudes from 
$-5$ ($10^2\, e^- {\rm \  total})$, to $-17$, and with colors that 
placed the stars along the fiducial cluster sequence, which followed the 
main sequence up the giant branch.  Stars brighter than about $-13.75$
are saturated in the deep images.  The exposure times for the deep
images for each cluster were chosen so that saturation would occur
above the sub-giant branch (SGB).  In the AS tests, when an added
star pushed a pixel above the saturation limit, we treated that pixel
as saturated in our finding procedure, but we made no attempt to model
how the added charge would bleed up and down the columns.  Thus,
brighter than the SGB, the artificial-star tests should be treated
more qualitatively than quantitatively.  Nonetheless, the qualitative
tests indicate that the completeness is essentially 100\% above the
SGB throughout almost all the clusters.  For the few clusters that 
are crowded and saturated at their centers, more sophisticated 
artificial-star tests may be required, but the fact that our data set 
has only one short exposure in each filter does limit what can be
done when the bright stars are crowded.

In order to sample the cluster radii evenly, we inserted the stars with 
a spatial density that was flat within the core, and declined as $r^{-1}$ 
outside of the core.  In this way, we performed the same number of tests 
in each radial bin.  Our standard artificial-star run had about $10^5$ stars 
and will be made available along with the real-star run for each cluster 
when we release the catalog.

\subsection{Using the artificial-star tests}
\label{ss.using_astests}
\begin{figure}
\plotone{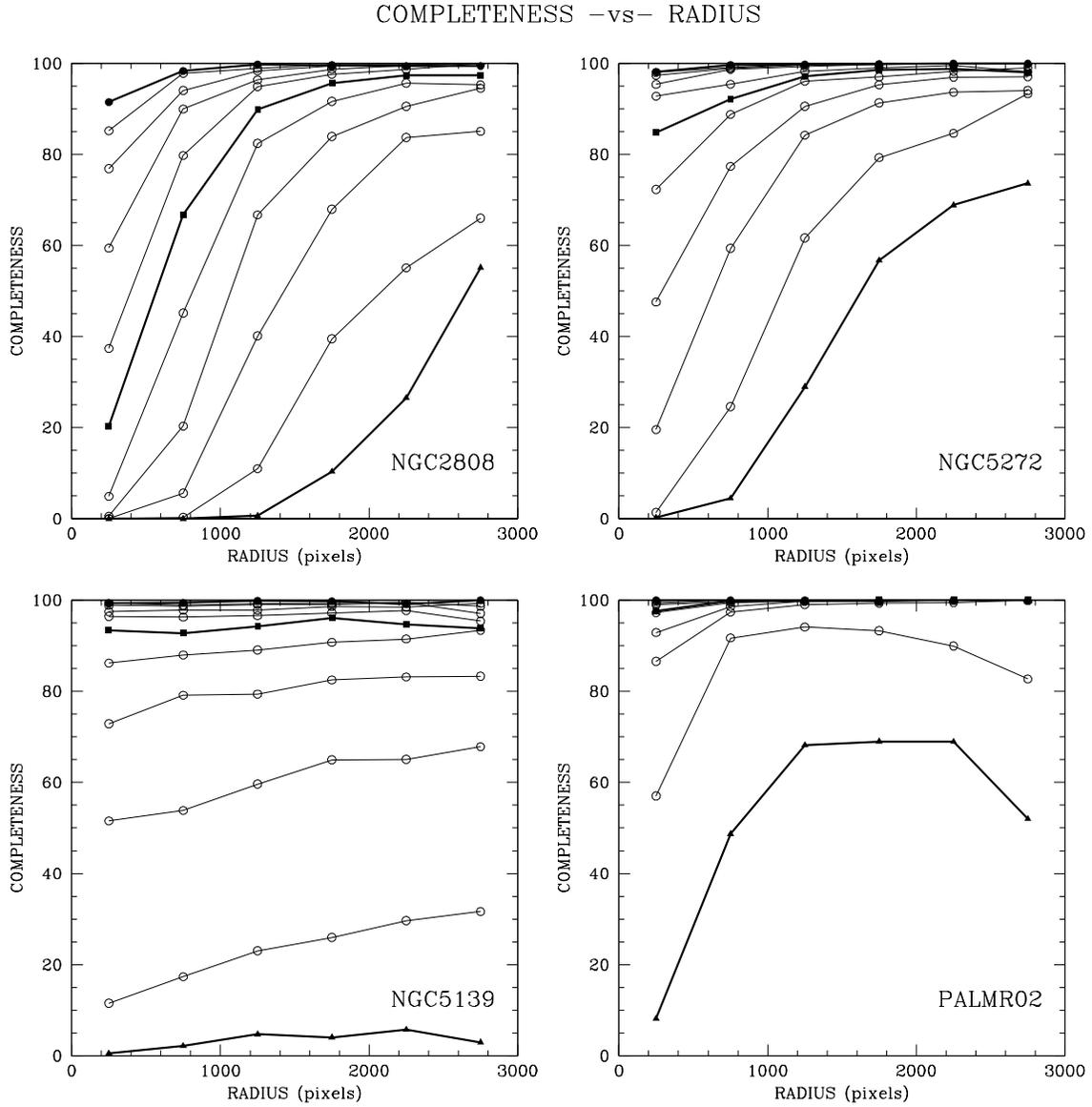}
\caption{The completeness fraction as a function of radius for four clusters.
         The lines show the completeness for bins 1.0 magnitude tall
         centered on {\tt mv} = $-5$ through $-15$.  The faintest bin, 
         at {\tt mv}=$-5$, is shown as a heavy line with filled-triangle 
         symbols.  The middle bin, at {\tt mv}=$-10$, is shown as a heavy 
         line with filled-square symbols.  The brightest bin, at 
         {\tt mv}=$-15$, is shown as a heavy line with filled circles.  
         The cluster main-sequence turn-off is typically at instrumental 
         magnitude $-12.5$.  
         \label{fig04}}
\end{figure}

The most obvious use of artificial-star tests is to assess completeness.
Figure~\ref{fig04} shows the completeness fractions as a function of
radius for four clusters in our sample.  NGC~2808 has a very crowded core,
and even stars near the turnoff (F606W $\sim\!-13$, in instrumental 
magnitudes) have moderately low completeness in the core.  NGC~5139 
($\omega$ Cen) has moderate crowding, but a very broad core, and so 
the completeness does not vary much with radius within our field.  
In NGC~5272, the completeness is almost 100\% for the brighter stars in 
the core, but fainter stars are lost there.  Finally, in the sparse 
Palomar 2 the completeness is almost 100\% everywhere for all but the 
very faintest stars.  

Most symbols in Figure~\ref{fig04} represent about 2000 AS tests,
so they should be accurate to about 2\%.  However, because the field is
square, the outer two bins contain fewer stars and should have
errors of 3\% and 7\%, respectively.  Also, the bottom curve
(for {\tt mv}=$-5$) contains only half as many stars as the others,
since the stars were inserted with a flat LF between $-5$ and $-17$.
Thus, the turndown for the faintest and furthest points in NGC~5139
and Pal 2 can be traced to small-number statistics.

Artificial-star tests can also be used to tell us about photometric
biases in the sample.  Some fraction of sources in the field are
superpositions of two stars that happen to lie nearly along the same
line of sight.  Sometimes, if the stars are not too close to one
another, the two can be disentangled by means of our multiple-pass
finding.  Other times, the quality-of-fit parameter can help to identify
blended stars that had a broadened profile, yet were too close to
separate.  Nonetheless, some superpositions are hard to identify and
will masquerade as photometric binaries.  The artificial-star tests can
be used to evaluate directly the contributions from these various
kinds of blends.


\section{PHOTOMETRIC ERRORS}
\label{s.PHOTO_ERRORS}
In \S~\ref{s.OVERVIEW} we made the point that different scientific
objectives are sensitive to different kinds of photometric errors.
Unfortunately, it is hard to come up with a single number to characterize
the photometric error for each star.  When we combined the
independent measurements for each star in \S~\ref{ss.measurement}, the
agreement among the independent measurements gave us some handle on the
measurement errors ($\sigma_V$ and $\sigma_I$).  However, there are some
systematic errors that cannot be detected in this way.  For instance, a
particular star will be found in the same place relative to the same
neighbors in all the exposures, so any error related to that crowding
will be the same for all measurements, and it will not show up in the
r.m.s.\ deviation, There are two main things that prevented us from
measuring each star as well as the r.m.s.\ errors would imply: the
presence of other stars and errors in the PSF.  In this section, we
discuss ways to identify and mitigate these sources of error.
 
\subsection{Errors related to crowding}
\label{ss.params}
The first way the magnitudes of a star can be compromised is by the presence 
of neighbors.  Thanks to our multiple-pass finding approach, we were able 
to find essentially any star that a careful human could find.  This enabled 
us to subtract off a good model of the neighbors of each star before we
measured the star itself.  This certainly improved our photometry, but 
neighbor subtraction can never be done perfectly, and it is invariably 
the case that isolated stars are measured better than stars with near 
neighbors.

In the course of computing the four basic parameters for each star (the 
$x$ and $y$ positions and $V$ and $I$ fluxes), we also came up with several 
additional diagnostic parameters that can be used to tell us how well each 
star was measured.  The most useful of these are:  
(1) $\sigma_V$ and $\sigma_I$, the r.m.s.\ deviation of the independent 
    flux measurements made in the different exposures, 
(2) $q_V$ and $q_I$, derived from the absolute value of the residuals of
    the PSF fit for each star (scaled by the flux),
(3) $o_V$ and $o_I$, the amount of flux in the aperture from neighboring 
    stars relative to the star's own flux, and
(4) $n_V$ and $n_I$, the number of images in which the star was found.

\begin{figure}
\plotone{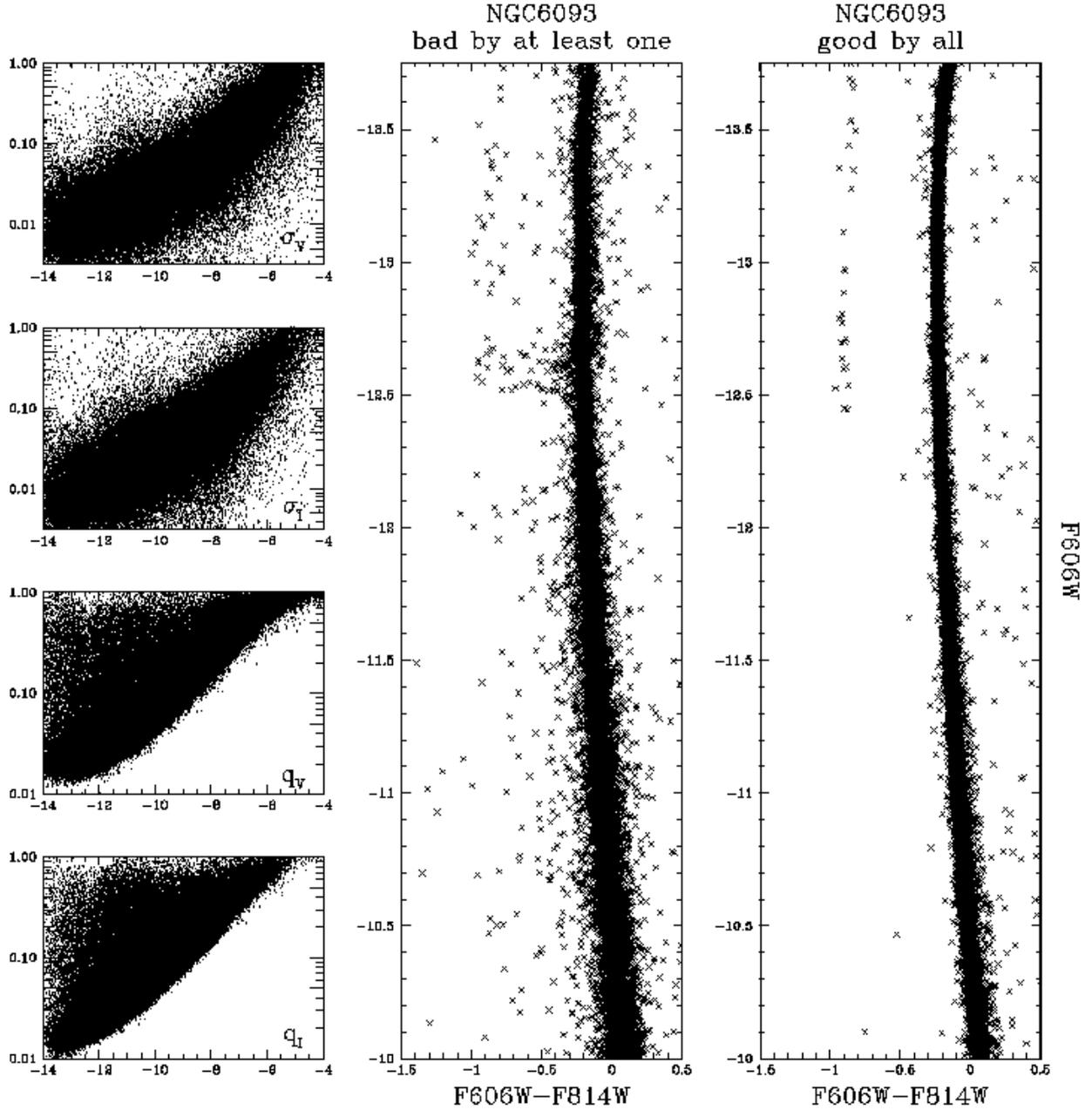}
\caption{In the left panels, we show from top to bottom $\sigma_V$, 
         $\sigma_I$, $q_V$, and $q_I$ as a function of instrumental 
         magnitude for
         the stars in NGC~6093.  The lines delineate the well-measured
         stars (those below the lines) from those that are less well 
         measured (above the lines).  The CMD in the middle panel shows 
         those stars that fell above the line in at least one of the four
         plots.  The CMD on the right shows the stars that appear to be well 
         measured according to all the parameters.
         \label{fig05}}
\end{figure}

These additional parameters can be used in two ways.  One way to use
them is on a star-by-star basis.  If there is a particular star of
interest in an unusual place in the CMD, then we can compare its
measurement parameters against those of stars of similar brightness
nearby to see if there may issues that might explain the photometric
peculiarities of the star.  Another way to make use of the additional
parameters is to identify a subset of stars that are more likely to be
better measured.  The left panels of Figure~\ref{fig05} show the trends
for the quality-of-fit and $\sigma$ parameters as a function of
magnitude for NGC~6093.  In each plot there is a locus of well-measured
stars near the bottom, and a more distended distribution of stars with
larger errors.  We drew in discrimination lines by eye to separate the
stars that were clearly poorly measured from those that were close to
the well-measured distribution.  A star had to be above the line in only
one of the four plots to be considered suspect.  The selections we have
made put about half the stars into the well-measured sample and half
into the more suspect sample.  On the right, we show CMDs for the two
samples.  It is clear that many stars that have photometry which places
them off the main sequence in the CMD also have larger internal errors
and/or poorer PSF fits.  This is the case both for stars well off the
sequence and for stars that are just a little off the sequence.  (The
sequence is much broader in the left CMD.)

\begin{figure}
\plotone{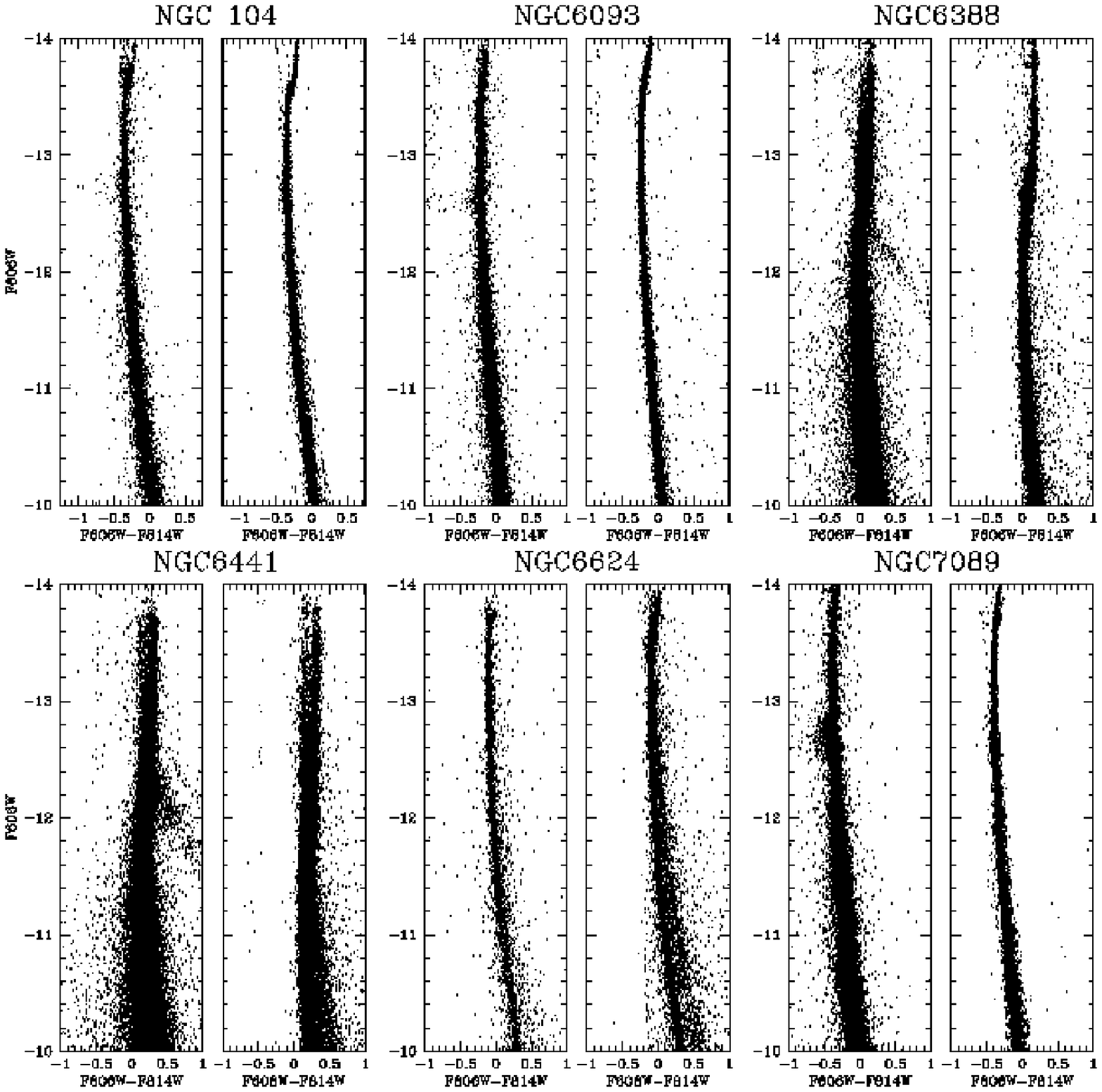}
\caption{We show the low-quality {\it (left)} and 
         high-quality {\it (right)} samples 
         for six selected clusters.  In NGC~6441 and NGC~6624, we see that 
         the second sequence is not an artifact of the photometry but 
         represents real, well-measured stars---likely a young, foreground 
         population.  
         \label{fig06}}
\end{figure}

Figure~\ref{fig06} shows the same selection strategy for six different 
clusters, with a variety of central concentrations.  For all the clusters, 
the quality-selection algorithm from the previous figure is able to identify 
stars that are not measured well.  We note that in crowded centers there 
is often a tuft of poorly measured stars at around F606W $\sim -12.5$ 
and F814W $\sim -12.5$ (the diagonal tufts in NGC~6388 and NGC~6441).  
We have visually inspected these stars in the images and found that these 
are stars near the crowded centers of clusters with nearby saturated 
neighbors that have bled into the star's aperture in one of the filters.  
Our modeling of the neighbors was not able to simulate such complicated 
artifacts, therefore a small number of stars suffered unavoidable 
contamination.  Thankfully, these stars can be identified by their large 
photometric errors.
 
Despite these clear improvements in the diagrams, the quality parameters 
should not be thought of as a panacea.  Imposing quality cuts on the data 
often implicitly imposes other selections as well.  For instance, stars that 
are more isolated are often better measured, so the quality cuts naturally 
select for stars in the less crowed outskirts of the clusters.  If the 
scientific goal is to study a feature in the CMD that should have no radial 
dependence (such as the turnoff morphology), then this will not affect the 
science.  But if the goal is to study blue stragglers or binaries, 
then any radial correlation between these populations and the quality 
parameters may well produce a biased sample.  An examination of the quality 
parameters as a function of radius could mitigate these selection effects.  

\subsection{PSF-related errors}
\label{ss.psf_errors}

The other kind of non-random error that can affect our photometry comes from
the PSF itself.  Ideally, we would like to measure each star with a large 
fitting radius or aperture (e.g., $\sim5$ pixels radius), so that our flux 
measurement for each star would have as little sensitivity as possible 
to the details of the PSF model.  Unfortunately, almost all of the stars 
in our fields have neighbors within this radius, and it would be very 
difficult to disentangle the light from overlapping star images over
such a large area.  It was obviously necessary to use a smaller fitting 
region in order to focus on the most relevant pixels for each star.  
Our standard fitting aperture was 5$\times$5 pixels, corresponding to 
a radius of $\sim2.5$ pixels.  When there was crowding, we often had 
to focus even more on the PSF core (see \S~\ref{ss.measurement}).  This 
necessary focus on the central regions of the PSF made us particularly 
vulnerable to any variations in the PSF that affected what fraction of 
light fell within the adopted fitting radius. 
 
To understand how PSF variation may have affected our photometry, it is 
important to consider how the WFC PSF can vary with position or with 
time.  Even if the PSF were perfectly constant over time, it would 
still have a different shape in different places on the detector due both
to distortion and to spatial variations in the chip's charge-diffusion 
properties caused by variations in chip thickness (Krist 2005).  On account
of both of these effects, the fraction of light in the central pixel of 
the F606W PSF varies from 18\% to 22\% from location to location on the 
detector.  If this is not accounted for, then fluxes measured by 
core-fitting can vary by up to 10\% (0.1 magnitude).  On top of this, 
spacecraft breathing can introduce an additional 5\% variation in the PSF 
core intensity.  To deal with these variations, our PSF model had a 
temporally constant component that varied with position, and a spatially 
constant component that accounted for how the PSF in each exposure differed 
from the library PSF.

Our two-component PSF model did a good job generating an appropriate 
PSF for each star in each exposure, but the model is not perfect.  
Unfortunately, when the telescope changes focus, the PSF does not change 
in exactly the same way everywhere on the detector, and there are 
residual spatially dependent variations of a few percent in the fraction 
of light in the core.  We considered constructing more elaborate PSF 
models, but there were simply not enough bright, isolated stars in 
these fields to allow us to solve for an array of corrections to the 
library PSF for each exposure.  To improve the PSF this way, we would have had 
to measure the PSF profile out to at least 5 pixels for a large number 
of stars distributed throughout the field.  The centers of most of our 
clusters were simply too crowded to permit us to model the PSF's spatial 
variation empirically.  Thus, there is a limit to how well we can know 
the PSF in each exposure, and this uncertainty naturally impacts our 
ability to measure accurate fluxes for the stars.  It is interesting 
to note that the same crowding that prevents us from using large 
apertures when we measure stars also prevents us from measuring much 
more than the core of the PSF in the centers of clusters.  
This further limits the accuracy of our measurements.  

The main effect that unmodelable PSF variations have on our photometry
is to introduce a slight shift in the photometric zero point as a
function of the star's location in the field.  On average this shift is
zero (thanks to the spatially-constant-adjustment part to the PSF), but
the trend with position can be as large as $\pm$0.02 magnitude.  These
small systematic errors will not be important at all for
luminosity-function-type analyses, where stars are counted in wide bins.
But the errors can be important for high-precision analyses of the
intrinsic width of CMD sequences or for studies of turnoff morphology.
In general, the PSF variation affects the F606W and F814W filters
differently, so the most obvious manifestation of this systematic error
is a slight shift in the color of the cluster sequence as a function of
location in the field.  This variation, in fact, is very hard to
distinguish from a variation in reddening with position, which
is certainly present in many of the clusters.

\begin{figure}
\plotone{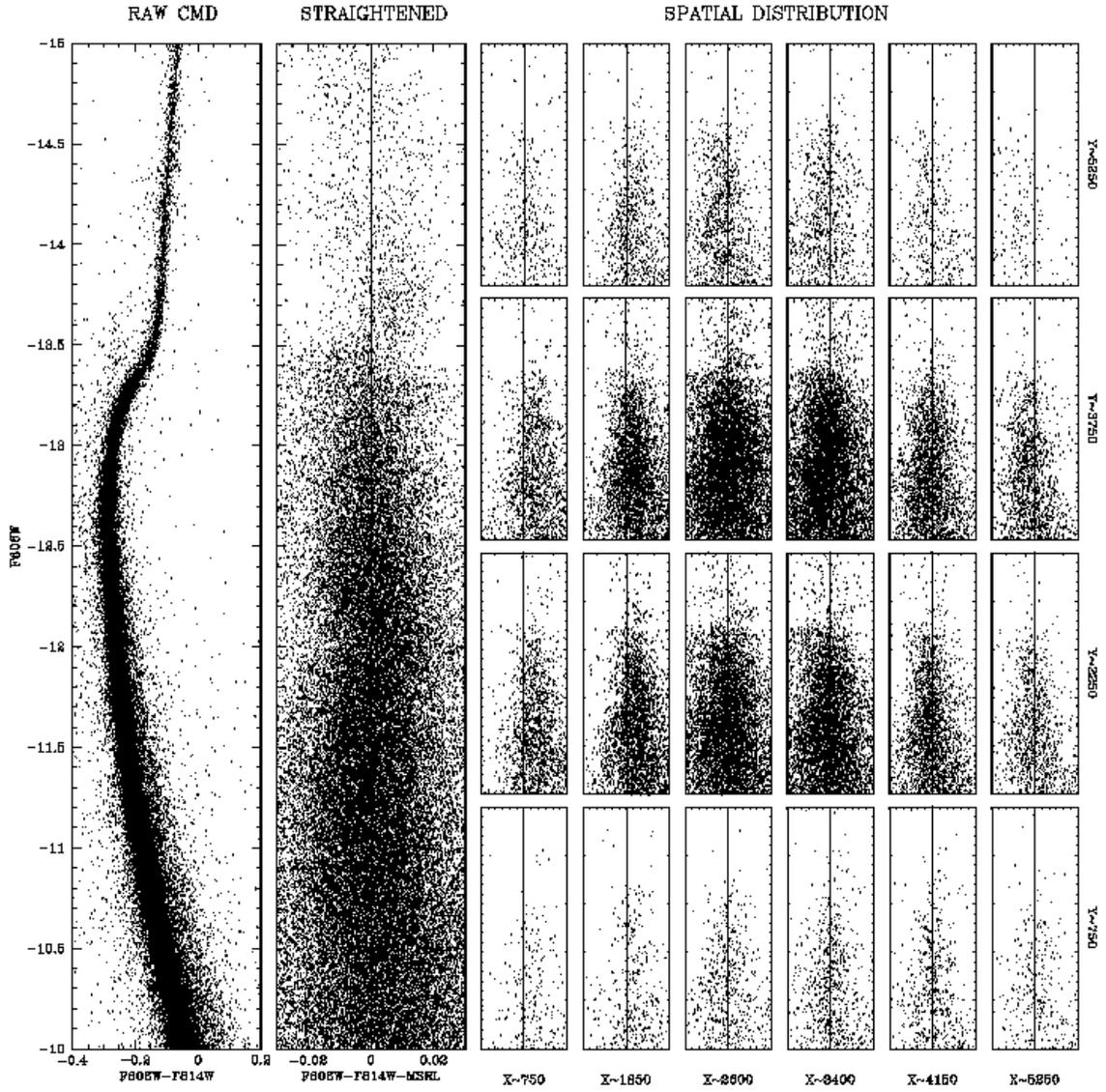}
\caption{{\it Left}: CMD for NGC~5272.  {\it Next panel}: CMD
         straightened by MSRL.  {\it Right set of panels}: the
         straightened sequence for an array of locations on the
         detector.
         \label{fig07}}
\end{figure}

In an effort to examine these color residuals, we first modeled the 
main-sequence ridge line (MSRL), as in the left panel of Figure~\ref{fig07} 
by tabulating the observed F606W$-$F814W color as a function of F606W 
magnitude.  We next subtracted from each star's observed color the MSRL 
color appropriate for its F606W magnitude.  This gave us a vertically 
straightened sequence (next panel of Fig.~\ref{fig07}), with a color 
residual for each star.   In the right array of panels in Figure 7, we 
examine the location of the observed sequence relative to the MSRL for 
different places within the field.  We see that in some places the 
cluster sequence systematically lies a little to the red or to the blue 
of the average MSRL.

\begin{figure}
\plotone{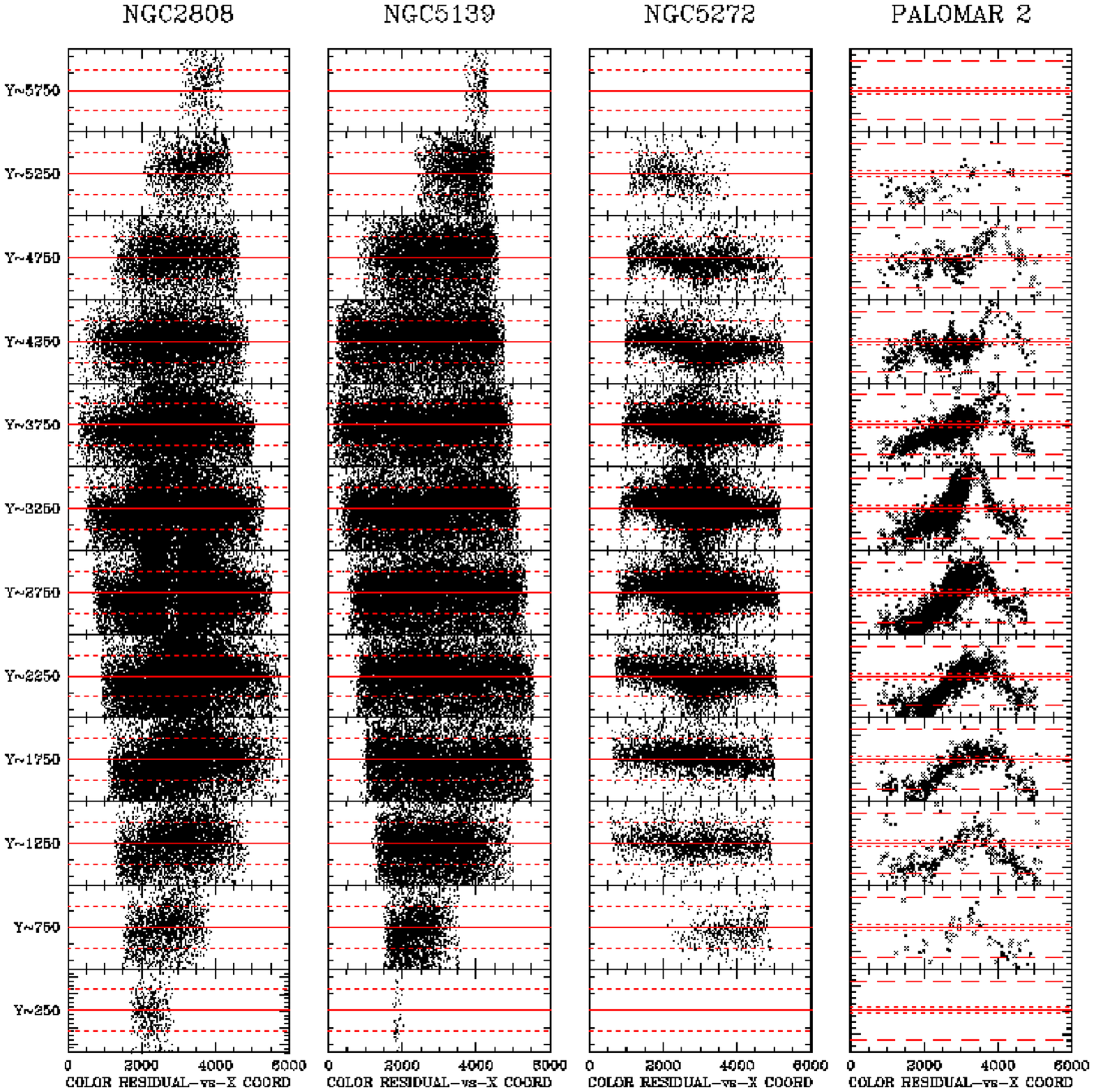}
\caption{The spatial dependence of the color residuals for four clusters.  
         We divide the 6000$\times$6000 field for each cluster into 
         12 horizontal slices, each 500 pixels tall in $y$ (the center is
         marked on the right).  Within each panel, we show the color residual 
         from the MSRL as a function of $x$ coordinate.  In each of the
         first three panels the dotted lines represent a color
         difference of $\pm0.025$ mag.  In the rightmost panel
         the dashed lines correspond to $\pm0.25$ mag.
         \label{fig08}}
\end{figure}

In Figure~\ref{fig08} we plot these color residuals for four clusters as 
a function of location on the chip.  For the first three, we see 
systematic residuals of $\pm$0.01 magnitude or so.  We know that these 
errors are often related to the PSF because when we have explicitly 
measured bright stars with larger apertures, the systematic trends were 
reduced (even though the spread about the MSRL is often greater, because of 
the stray light that enters a larger aperture).  These systematic errors 
may seem quite small, but from the r.m.s.\ spread of the independent
and artificial-star tests we would expect color errors of less than 
0.005 magnitude for each well-exposed star, so the systematic trends do 
limit how well we can evaluate the intrinsic width of the sequence for 
each cluster.  

The cluster on the right (Pal 2) exhibits color residuals of
$\sim0.20$ mag, more than ten times those for our typical cluster.  
These residuals are due to variable reddening for this low-latitude 
cluster and are largely unrelated to the PSF.  Reddening has a similar 
effect to that of the PSF-related shifts, except that stars affected by reddening
should be shifted along the reddening vector, while PSF-related shifts do 
not necessarily have their $V$ and $I$ shifts correlated.

One way to mitigate the color effect is to introduce an array of empirical 
corrections across the field and adjust the color for each star according 
to this table.  This procedure does tend to tighten up the CMD and allows
us to see more structure (see Milone et al.\ 2007 for a study of 
the NGC~1851 CMD), but it is hard to do highly accurate work this 
way. 

\clearpage


\section{THE FINAL CATALOG}
\label{s.CATALOG}

The procedures described thus far have produced instrumental magnitudes,
and positions in an adopted reference frame for each cluster.  For
our final catalog, however, we need to put the magnitudes onto correct
zero points, and give positions in an absolute frame.  In addition,
improvements are needed in the photometry of the saturated stars, and
corrections must be made for the effects of CTE.

\subsection{Improving the brightest stars}
\label{ss.satphot}
In designing this project, we chose the length of the short exposures in
each cluster in such a way that the horizontal branch would be well exposed 
but not saturated.  Even though the brighter RGB stars were also of interest, 
it was not efficient to take more than one short exposure for each cluster.  
The automated finding program discussed in \S\ \ref{s.KSYNC} did find the 
saturated stars, and it measured a flux for each one by fitting the wings of 
the PSF to the unsaturated pixels; but such measurements tend to have 
large errors, both random and systematic.

There is a better way of measuring the saturated stars.
Gilliland (2004, G04) has found that when a star saturates in the WFC,
its electrons bleed into other pixels, but the total number of electrons
due to that star is conserved.  If the gain is set to 2, then this
information is preserved in the {\tt flt} image.  We were able to verify
that the procedure recommended in G04 works for our images, by using it
on the stars that are saturated in our long-exposure images, and
comparing the resulting fluxes with the accurate fluxes that we had
measured for those same stars in our short exposures.  The technique
that we used was to measure each star in an aperture of 5-pixel radius
and include in addition the contiguous saturated pixels that had bled
even farther.  We found that the fluxes that we measured in this way
agreed well with those measured from the unsaturated images in the short
exposures.  Thus we felt confident in our use of the G04 technique to
measure the saturated images in the short exposures, and used these
measurements for our final instrumental magnitudes of those stars.

\begin{figure}
\plotone{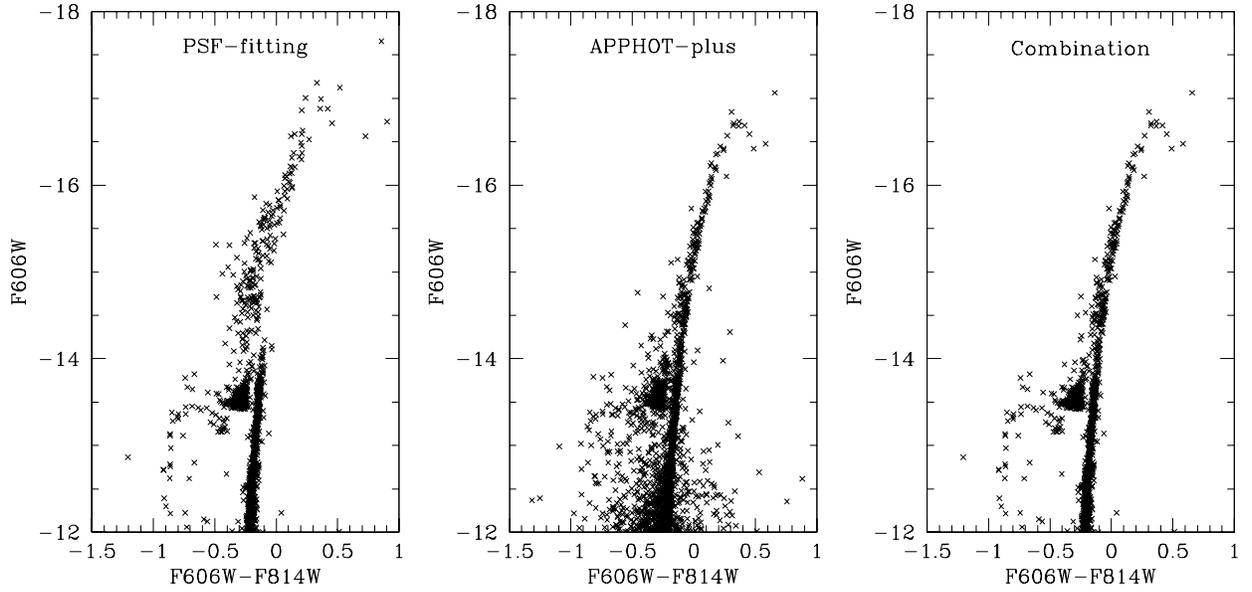}
\caption{{\it (Left)} The upper part of the CMD for NGC~2808 for stars
         measured with the wings of the PSF.  {\it(Center)} The same
         stars, but with the aperture-based approach.  {\it (Right)}
         Combination of the two:\ PSF-fitting for unsaturated stars, and
         the aperture-based approach for the saturated stars.
         Magnitudes are instrumental; saturation sets in at around
         $-13.75$.
         \label{fig09}}
\end{figure}

Figure~\ref{fig09} shows a comparison between the CMD obtained from
PSF-fitting and the one obtained from the G04 approach.  The
improvement in the upper parts of the CMD is dramatic, both in the
continuity of the sequences and in the photometric spread.  Note that
towards the bottom of the middle plot the photometric errors increase
significantly.  This is because the 5-pixel aperture often includes more
than just the target star, even for these bright giant-branch stars.
PSF-fitting is clearly much better than aperture photometry when stars
are not saturated, since most of our accuracy comes from the few central
pixels, with their high signal-to-noise ratio.  The final photometry uses
the better measurement for each star:  for stars that are unsaturated in
the short exposures, we use the PSF-fit result, but for saturated stars
we substitute the aperture-based result.

We became aware of the G04 approach only after a large number of the
clusters had already been measured.  If we had known of it from the
beginning, we would have incorporated it directly into our procedures
instead of making it a separate post-processing step.

\clearpage

\subsection{CTE corrections}
\label{ss.cte}
The background in many of our short exposures is low enough to raise
concerns about the impact of CTE effects on our photometry.  The
standard corrections for CTE effects are provided for aperture
photometry with several aperture sizes, by Riess \& Mack (2004,
RM04).  Since our photometry comes from PSF fitting to the inner
5$\times$5 pixels rather than from aperture photometry, it is unclear
which aperture is the most appropriate match to our measurements. In the
light of this ambiguity, we proceeded as follows:

We used Eq.\ 2 of RM04,
$$
    YCTE = 10^A \times SKY^B \times FLUX^C 
                             \times {y_{\rm readout}\over{2048}} 
                             \times {(MJD - 52333)\over{365}},
$$
to estimate the CTE correction for each star, given the local 
sky background, the $y$-position of the star in the {\tt flt} images, 
the Modified Julian Date (MJD) of the observation, and the flux 
of each star as determined from the PSF magnitude.  The quantity 
$y_{\rm readout}$ is the number of $y$ shifts experienced by the pixel;
it is simply $y$ for the bottom chip and 2049-$y$ for the top chip.

\begin{figure}
\plotone{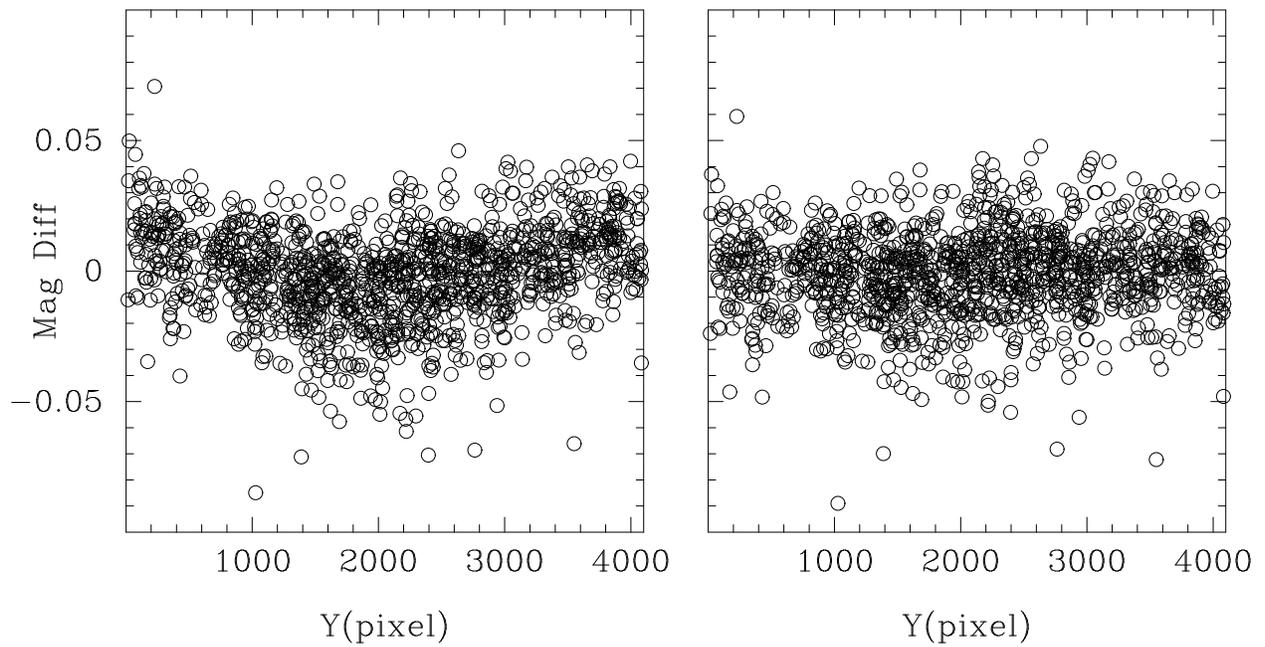}
\caption{The left panel shows the difference between the instrumental 
         magnitudes on the long-exposure frames and the short exposures, 
         for stars in common, as a function of y-position, before the 
         application of the CTE correction for the F606W observations of
         NGC~6809.  The right panel shows the same stars after the 
         CTE correction has been applied.
         \label{fig10}}
\end{figure}

RM04 provide values of the exponents $A$, $B$, and $C$ for various sizes
of the photometric aperture.  We chose the values for a 5-pixel
aperture, and made those corrections, typically $\sim0.02$ mag, to our
PSF photometry.  We then compared the short- and long-exposure
photometry for the same stars, after both had been CTE corrected.  We
then examined the the magnitude differences between the short- and
long-exposure photometry, and for almost all clusters the mean
difference was zero, with no significant trend as a function of the
input parameters $y$-position, sky background, and stellar flux.  (See
Figure~\ref{fig10} for an example.)  In a few cases there was a
systematic variation as a function of $y$-position; for these clusters
we adopted an aperture size of 7 pixels in the calculation of the CTE
corrections, and that eliminated the trend.

\clearpage

\subsection{Photometric calibration}
\label{ss.photo_calib}

Thus far we have kept our photometry in instrumental magnitudes, because
of their simple relation to counted electrons.  (As stated in \S\
\ref{ss.ref_frame}, instrumental magnitudes are simply $-2.5\log N$,
where $N$ is the number of counted electrons in the first deep {\tt flt}
image.)  We now need to put our magnitudes on a correct zero point.

Unfortunately, our instrumental magnitudes refer to {\tt flt} images,
but the zero point definitions provided by STScI refer to {\tt drz}
images.  We therefore measured a few dozen isolated bright stars in the
the {\tt drz} images, using the procedure detailed in Bedin et al.\,
(2005).  We then used the encircled-energy corrections and the zero
points given by Sirianni et al.\, (2005, S05) to arrive at calibrated
VEGAMAG photometry:
$$ 
m_{\rm filter} =
 - 2.5 ~ \log_{10} \frac{I_{\rm e^-}}{exptime}
 + Zp^{\rm filter}
 - \Delta m_{AP_{0.\!\!^{\prime\prime}\!5}-AP_{\infty}}^{\rm filter} 
 - \Delta m_{PSF-AP_{0.\!\!^{\prime\prime}\!5}}^{\rm filter}, 
$$
where ``filter'' refers to either F606W or F814W.  The first term 
on the right refers to the PSF-fitting photometry in the {\tt flt} images, 
the second term is the zeropoint (from S05's Table 10), and the third term 
is the correction from the 0\secspt5 aperture to the nominally infinite 
aperture (from S05's Table 5).  The final term must be measured empirically 
as the difference between our PSF-fitting photometry and the 
0\secspt5-aperture photometry in the {\tt drz} image.  This is 
typically close to zero, since our PSFs have been normalized to have 
unit volume within a radius of 10 {\tt flt} pixels.

\begin{figure}
\plotone{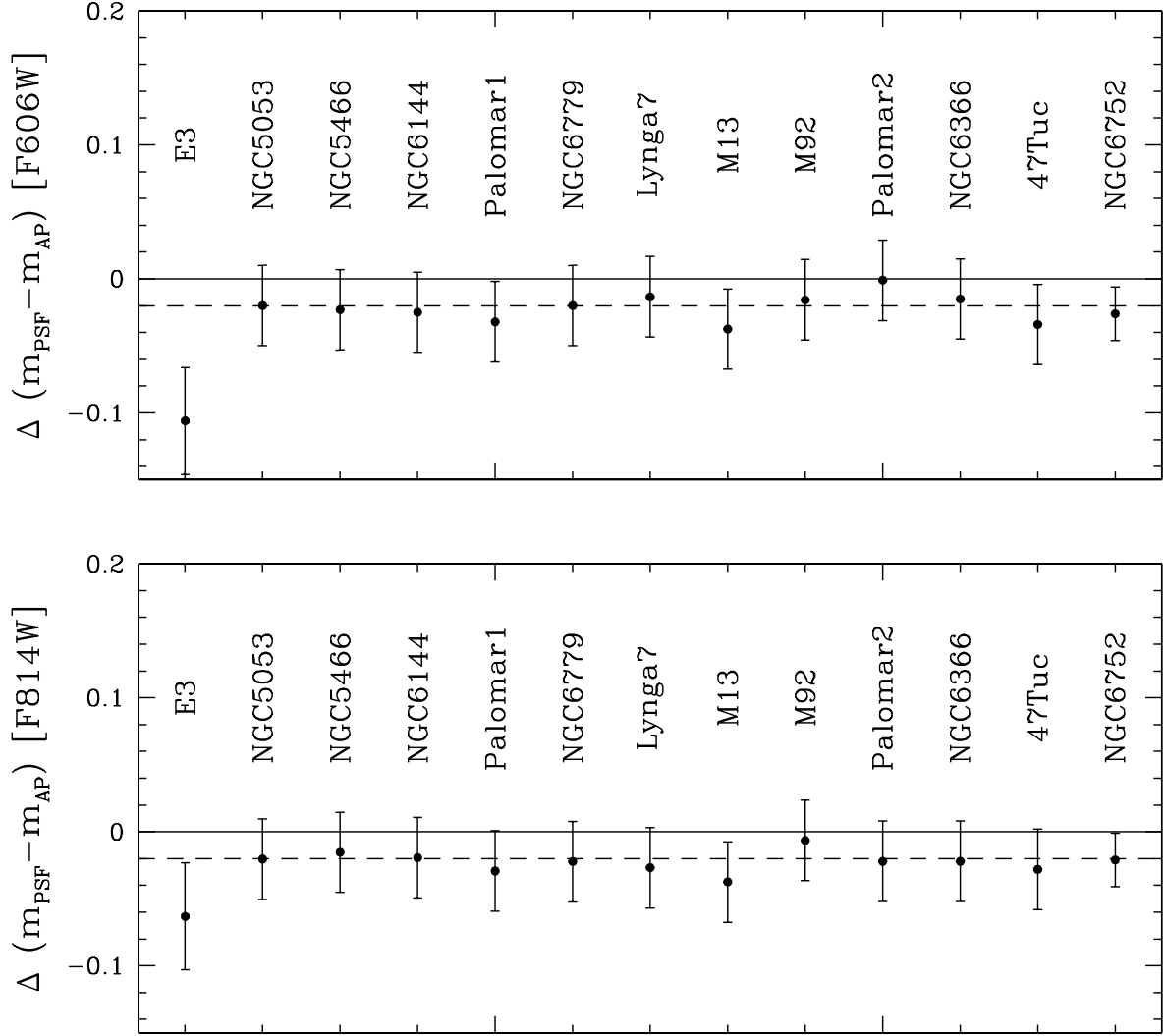}
\caption{The zero-point difference between the PSF-fitting photometry on
         the {\tt flt} images and the 0\secspt5-aperture on the {\tt
         drz} images, determined empirically for several clusters.  The
         error bars indicate the range of stars measured for that cluster, 
         reflecting both random errors and possible systematic errors with 
         position.  The dashed line shows the $-0.02$ value adopted as the 
         average.  The PSF for the E3 images was observed to be more out of
         focus than for any other cluster.  Also, the E3 field is sparse,
         which makes it hard to improve the PSF model with an accurate
         perturbation PSF.
         \label{fig11}}
\end{figure}

Figure~\ref{fig11} shows the {\tt flt}$-${\tt drz} term (the fourth
term) for several clusters for which it was easy to measure.  Several of
our clusters were so crowded---even in the outskirts---that we could not
find enough isolated, unsaturated stars to measure an uncontaminated
flux within the 0\secspt5 calibration radius.  Since the offset appears
to be constant (as it should be), we simply adopted the average value
over all the clusters ($-0.02$ magnitude).  We expect the absolute
calibration to be accurate to about 0.01 magnitude for the typical
cluster, but because of focus variations that affect the PSF (see
\S~\ref{ss.psf_errors}), the zero-point errors can approach 0.02
magnitude and can vary with position in the field.

\clearpage

\subsection{Absolute astrometric frame}
\label{ss.abs_astrom}

The reference frame we adopted for each cluster was based on the WCS
information that the reduction pipeline had placed in the header of the
{\tt drz} image.  (See \S\ \ref{ss.ref_frame}.)  We expect the absolute
astrometric zero point for this frame to be accurate only to
1--2\arcsec, since that is what can be expected from errors of the
absolute positions of HST's guide stars (Koekemoer et al.\, 2005).

To get zero points that were more accurate, we downloaded the 2MASS
point-source survey for for the region of each cluster, and found
between 40 and 1500 reference stars that we were able to match up with 
stars in our lists.

\begin{figure}
\plotone{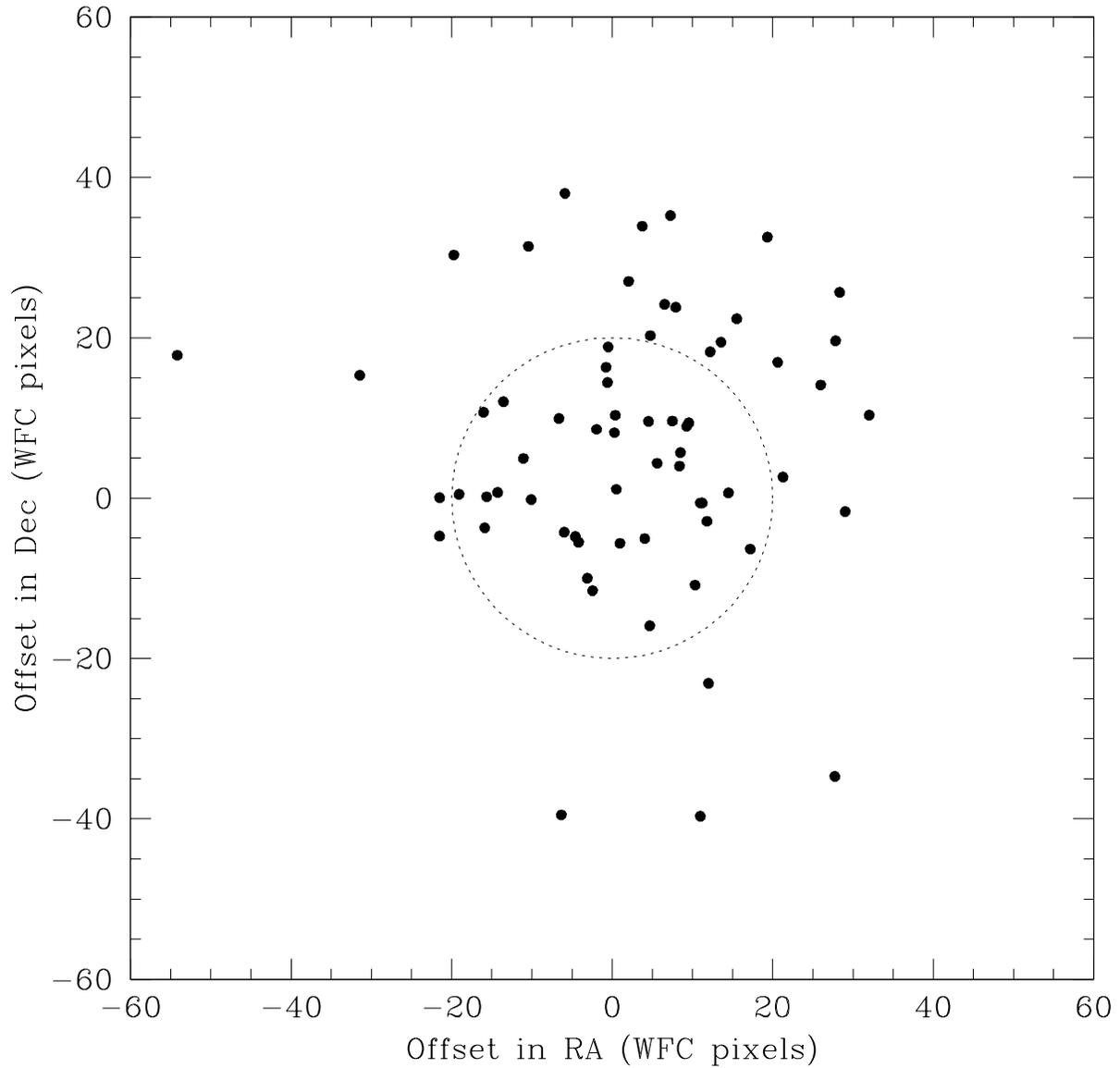}
\caption{Offsets between absolute positions constructed from the WCS header 
         of the {\tt drz}-frame and positions given in the 2MASS catalog.  
         Each point represents one cluster.  The dotted circle corresponds 
         to 1\arcsec\, (20 pixels).  Positions in our final catalog have 
         been shifted to agree with the 2MASS zero points.  
         \label{fig12}}
\end{figure}

We then compared our absolute positions against the absolute positions
of the same stars in the 2MASS catalog, and found that the two frames
were typically offset by $\sim$1.5\arcsec .  Figure~\ref{fig12} shows
the distribution of offsets for the ensemble of clusters.  The typical
shift is consistent with the expected astrometric accuracy of HST's
guide-star catalog.  Each measured shift came from averaging many tens of 
stars, each with a typical residual of 0\secspt15.  Thus our final absolute
frame for each cluster should have an absolute accuracy much better than
this.  (The absolute accuracy of 2MASS positions is given as 15 mas in
Skrutskie et al.\, 2006.)  We adjusted the WCS header in each of our
stacked images (which will be included with the catalog), to reflect the
improved absolute frame. 

The relative positions of stars in our field should be much more
accurate than their absolute zero point (15 mas corresponds to 0.3 pixel).
The non-linear part of the WFC distortion solution is accurate to better
than 0.01 pixel (0.5 mas) in a global sense (see Anderson 2005), which
is about the random accuracy with which we can measure a bright star in
a single exposure.  Recently, it has been discovered that the linear
terms of the distortion solution have been changing slowly over time
(see Anderson 2007).  Since our reference frames were based on the {\tt
drz} images (which had not been corrected for this effect), our final
frames contain an error of about 0.3 pixel in the off-axis linear terms.
Users are therefore cautioned to adopt general 6-parameter linear
transformations when relating our frame to other frames.  If such
transformations are made, our positions should be globally accurate to
0.01 pixel across the field.

\subsection{The main catalog}
\label{ss.final_cat}
Our entire catalog contains over 6 million stars for 65 clusters, with a
median number of 67,000 stars per cluster.  Our procedures generated a
large amount of information for each star in each cluster, but most
users will need only the high-level data for each star.  So for each
cluster we produced a single file called {\tt NGCXXXX.RDVIQ.cal}, which
has one line for each star found.  The columns give reference-frame
position, calibrated (i.e., zero-pointed) magnitudes, errors, calibrated
RA and Dec, and some general measurement-quality information.  The
column by column description for this file is given in
Table~\ref{tab04}.

\begin{table}
\begin{center}
\caption{Information in the {\tt .RDVIQ.cal} file. \bigskip }
\label{tab04}
\begin{tabular}{r|cl}
\hline 
\multicolumn{1}{ r|}{Col} &
\multicolumn{1}{ c }{Name} &
\multicolumn{1}{ l }{Explanation} \\
\hline 
 1 & {\tt ID}     & ID number for each star (same as line number) \\
\hline 
 2 & {\tt xref}   & average ref-frame $x$ position  \\
 3 & {\tt yref}   & average ref-frame $y$ position   \\
\hline 
 4 & ${\tt V}_{\tt VEGA}$   & calibrated F606W magnitude (in the VEGA-mag system)     \\
 5 & $\sigma_{\tt V}$       & RMS error in F606W photometry   \\
 6 & ${\tt VI}_{\tt VEGA}$  & calibrated F606W-F814W color    \\
 7 & $\sigma_{\tt VI}$      & RMS error in color              \\
 8 & ${\tt I}_{\tt VEGA}$   & calibrated F814W magnitude      \\
 9 & $\sigma_{\tt I}$       & RMS error in F606W photometry   \\
\hline 
10 & ${\tt V}_{\tt gnd}$    & photometry calibrated to ground-based $V$ \\
11 & ${\tt I}_{\tt gnd}$    & photometry calibrated to ground-based $I$ \\
\hline 
12 & ${\tt N}_{\tt V}$      & number of $V$ exposures star was found in \\
13 & ${\tt N}_{\tt I}$      & number of $I$ exposures star was found in \\
\hline 
14 & {\tt wv }              & source of $F606W$ photometry \\
   &                        & 1=unsaturated in deep ; 2=unsaturated in short;\\
   &                        & 3=saturated in short ; 4=saturated in deep    \\
15 & {\tt wi }              & source of $F814W$ photometry \\
\hline 
16 & {\tt ov }  & fraction of light in F606W aperture due to neighbors \\
17 & {\tt oi }  & fraction of light in F814W aperture due to neighbors \\
18 & {\tt qv }  & quality of F606W PSF-fit (smaller is better) \\
19 & {\tt qi }  & quality of F814W PSF-fit \\
\hline 
20 & {\tt RA }  & Right ascension for the star, in degrees \\
21 & {\tt Dec } & Declination, in degrees \\
\hline 
\end{tabular}
\end{center}
\end{table}


In addition, for each cluster we generated several auxiliary files, 
which contain the simultaneous-fit fluxes, the exposure-by-exposure 
photometry, and much more.  Finally, we also put together a similar set 
of files for the artificial-star tests, along with the list of input 
parameters ({\tt x\_in}, {\tt y\_in}, {\tt mv\_in}, and {\tt mi\_in}).
The stacked image in each color will be made available along with the 
catalog for each cluster.

\clearpage


\section{SUMMARY}
\label{s.SUMMARY}
The ACS Survey of Globular Clusters is the first truly uniform, deep
survey of the central regions of a large number of Galactic globular
clusters.  The observations for each of the 65 clusters were carefully
planned in order to provide even spatial coverage of  a 3$\times$3-arcminute 
region near the center of each cluster.  To make use of the uniformity 
of the observations, we developed a reduction strategy that would
reduce the data set for each cluster in an automated way, finding as
many stars as possible while at the same time minimizing the inclusion
of false detections.  The stars found were measured as accurately as
possible with the best available PSF models.

We adjusted the exposure times for individual clusters in such a way
that the final catalog of stars is largely complete down to
0.2\,$M_{\odot}$ for the less crowded clusters.  We hope that our nearly
definitive list of stars will make it easier for future researchers to
cross-identify stars in past and future cluster observations.

In addition to the catalog of real stars, we also constructed a standard
catalog of artificial-star tests for each cluster that can help assess
any incompleteness or photometric biases in the sample.  We plan to make
this catalog public in the near future with full access to the
photometric and astrometric data for each of the 65 clusters via the
world-wide web.

Even a cursory glance at the many CMDs in this survey shows that while 
the clusters all have the same general features, each cluster contains 
a unique population of stars, representative of its particular 
star-formation and dynamical history.  An early version of this catalog 
has already led to several papers, including:\
(1) a study of clusters with no previous HST observations 
    (Sarajedini et al.\ 2007);
(2) the creation of a set of stellar evolutionary tracks matching 
    our photometric system (Dotter {\ al.}\, 2007);
(3) population analysis of the M54/Sgr Color-Magnitude Diagram 
    (Siegel et al.\ 2007); and
(4) discovery of the multiple subgiant branch of NGC~1851 
    (Milone et al.\ 2007).

Additional papers are in preparation to study radial profiles, 
relative ages, cluster mass functions and mass segregation, the 
Sagittarius clusters, horizontal-branch morphology, the binary populations 
and their radial gradients, blue stragglers, internal proper motions, and 
dynamical families of clusters.
 
\acknowledgements

The Co-I's based in the United States acknowledge the support of STScI
grant GO-10775.  We thank the anonymous referee for thoughtful comments
that helped us make this more accessible to the community.

\references

\parindent -0.10in
\narrower

Anderson, J. \& King, I. R. 2006,  ACS/ISR 2006-01,  
   PSFs, Photometry, and Astrometry for the ACS/WFC

Anderson, J. 2007, ACS/ISR 2007-08,
   Variation of the Distortion Solution of the WFC.

Armandroff, T. E. 1989, AJ, 97, 375 

Bedin, L. R., Cassisi, S., Castelli, F., Piotto, G., Anderson, J., Salaris, M., Momany, Y., \& Pietrinferni, A.  2005, MNRAS, 357, 1038

Bedin, L. R., Piotto, G., King, I. R., \& Anderson, J. 2003, AJ, 126, 247 

Djorgovski, S., \& Meylan, G. 1993, Structure and Dynamics of Globular
   Clusters (eds.\ S.\ Djorgovski \& G.\ Meylan), ASPCS, vol.\ 50 (San
   Francisco:\ Astron.\ Soc.\ Pacific)

Djorgovski, S., \& King, I. R. 1986, ApJ, 305, L61

Dotter, A., Chaboyer, B., Jevremovi\`c D., Baron, E., Ferguson, J. W., Sarajedini, A., \& Anderson, J.  2007, AJ, 134, 376

Fruchter, A. S., \& Hook, R. N. 2002 PASP 114 792

Gilliland, R. 2004, ACS/ISR 2004-01, 
   CCD Gains, Full Well Depths, and Linearity up to and Beyond 
   Saturation

Harris, W. E. 1996, AJ, 112, 1487

Koekemoer, A. M., McLean, B., McMaster, M., \& Jenkner, H. 2005, ACS/ISR
   2005-06, Demonstration of a Significant Improvement in the
   Astrometric Accuracy of HST Data

Lee, Y. -W., Demarque, P., \& Zinn, R. 1994, ApJ, 423, 248

Milone, A. P., Villanova, S., Bedin, L. R., Piotto, G., Carraro, G., Anderson, J., King, I. R., \& Zaggia, S.  2006, A\&A, 456, 517

Milone, A. P., et al.  2007, ApJ, in press

Piotto. G., et al.  2002, A\&A, 391, 945

Riess, A., \& Mack, J. 2004.  ACS/ISR 2004-06, 
   Time Dependence of ACS CTE Corrections for Photometry and 
   Future Predictions

Rosenberg, A., Piotto, G., Saviane, I., \& Aparicio, A. 2000a, A\&AS, 144, 5

Rosenberg, A., Aparicio, A., Saviane, I., \& Piotto, G.  2000b, A\&AS, 145, 451

Sarajedini, A., et al. 2007, AJ, 133, 1658

Siegel, M. H., et al. 2007 ApJL 667 57

Skrutskie, M. F., et al. 2006, AJ, 131, 1163

Trager, S. C., King, I. R., \& Djorgovski, S.  1995, AJ, 109, 218

Zinn, R.  1980, ApJS, 42, 19

\clearpage

\clearpage

\end{document}